\documentclass[twocolumn,pra, aps,superscriptaddress,showpacs]{revtex4}

\usepackage{lineno}
  \usepackage{mathptmx}
\usepackage{subfigure}
\usepackage{sublabel}
\usepackage{dcolumn}
\usepackage{amsmath,amssymb}
\usepackage{bm}
\usepackage{bbm}
\usepackage{color}
\usepackage{overpic}
\usepackage{latexsym}
\usepackage{epstopdf}
\usepackage{color}
\usepackage[english]{babel}
\usepackage{latexsym}
\usepackage{stmaryrd}

\usepackage{psfrag,graphicx} 
\usepackage{epsf} 
\usepackage{subfigure} 
\usepackage{amsmath} 
\usepackage{amssymb} 
\usepackage{amsfonts}
\usepackage{bm}
\usepackage{natbib}
\usepackage{epstopdf}
\usepackage{appendix}

\definecolor{mygrey}{gray}{0.35}
\definecolor{myblue}{rgb}{0.2,0.2,0.8}
\definecolor{myzard}{cmyk}{0,0,0.05,0}
\definecolor{mywhite}{rgb}{1,1,1}
\definecolor{myred}{rgb}{1,0.,0.3}

\usepackage[colorlinks=true,citecolor=myblue,linkcolor=myred]{hyperref}
\def\be{\begin{equation}}
\def\ee{\end{equation}}
\def\ba{\begin{align}}
\def\enda{\end{align}}
\def\bi{\begin{itemize}}
\def\ei{\end{itemize}}

\newcommand{\sgn}{\operatorname{sgn}}

\def\dd{\mathord{\rm d}} 
 \def\ee{\mathord{\rm e}}
 
 \def\ii{\mathord{\rm i}}

\def\tr{\mathop{\rm Tr}}
\def\half{\textstyle\frac{1}{2}}

\def\fourth{\textstyle\frac{1}{4}}

\def\dd{\mathord{\rm d}} 
 \def\ee{\mathord{\rm e}}
 
 \def\ii{\mathord{\rm i}}

\def\tr{\mathop{\rm Tr}}
\def\half{\textstyle\frac{1}{2}}

\def\fourth{\textstyle\frac{1}{4}}

\renewcommand{\ii}{{\rm i}}
\renewcommand{\ee}{{\rm e}}

\def\beq{\begin{equation}}
\def\beq{\begin{equation}}
\def\eeq{\end{equation}}

 \newcommand{\ket}[1]{|#1\rangle}
 \newcommand{\bra}[1]{\langle #1|}

 \newcommand{\ketbradif}[2]{\ket{#1}\bra{#2}}
 \newcommand{\ketbra}[1]{\ketbradif {#1}{#1}}

\begin{document}

%\modulolinenumbers[1]
%\linenumbers

\title[Short Title]{Driven  Geometric Phase Gates with Trapped Ions }

\author{A. Lemmer}
\affiliation{Institut f\"ur Theoretische Physik, Albert-Einstein Alle 11,
Universit\"at Ulm, 89069 Ulm, Germany}

\author{A. Berm{u}dez}
\affiliation{Institut f\"ur Theoretische Physik, Albert-Einstein Alle 11,
Universit\"at Ulm, 89069 Ulm, Germany}

\author{M. B. Plenio}
\affiliation{Institut f\"ur Theoretische Physik, Albert-Einstein Alle 11,
Universit\"at Ulm, 89069 Ulm, Germany}

\begin{abstract}
We describe a hybrid laser-microwave scheme to implement two-qubit geometric phase gates in crystals of trapped ions. The proposed gates can attain errors below the fault-tolerance threshold in the presence of thermal, dephasing, laser-phase, and microwave-intensity noise. Moreover, our proposal is technically less demanding than previous schemes, since it  does not require a laser arrangement with interferometric stability. The laser beams are tuned close to a single vibrational sideband to entangle the qubits, while  strong microwave drivings provide the geometric character to the gate, and thus protect the qubits from these different sources of noise. A thorough analytic and numerical study of the performance of these gates in realistic noisy regimes is presented.

\end{abstract}

%\date{\today}
\maketitle

\section{Introduction}

Quantum information processing (QIP) holds the promise of solving certain computational tasks more efficiently than any classical device~\cite{nielsen_chuang}. This prospect has stimulated  an enormous technological effort, whereby  prototype quantum processors based on different technologies have already been developed~\cite{qc_review}. The promised supremacy of these  processors relies on the vast  parallelism available at the quantum realm, which in turn rests on the quantum superposition principle. Therefore, any quantum  processor must be well-isolated from environmental sources of noise, since these tend to degrade quantum superpositions through the phenomenon of decoherence. Additionally,  to benefit from the large parallelism, quantum processors should also allow for a very accurate manipulation of the information. Experimental imperfections in such manipulation will also conspire  to reduce the potential of QIP. 

Hence,  the design of QIP protocols that are {\it robust} to the most important sources of environmental and experimental noise, is considered to be a task of primary importance. The particular level of robustness required, which can be quantified by the maximal allowed error $\epsilon$ per step of the routine, is determined by the possibility of implementing protocols of quantum error correction that allow for fault-tolerant QIP~\cite{qec_preskill}. The so-called fault-tolerance threshold, $\epsilon_{\rm FT}$, typically depends on the dominant source of noise,  the particular error-correcting scheme, technological limitations of the experimental platform, and the particular purpose of the routine (i.e. initialization, manipulating information, or measurement). The manipulation of the information is argued to hold the most stringent thresholds~\cite{qc_knill}, and although some error-correcting schemes with thresholds as high as $\epsilon_{\rm FT_{1}}\sim 10^{-2}$ exist, it is commonly agreed that  reducing errors below  $\epsilon_{\rm FT_{2}}\sim 10^{-4}$ is a guiding principle for the development of quantum processors. Since the information is usually stored in the so-called qubits (i.e. two-level systems), these routines can be divided in prescribed sequences of one- and two-qubit gates allowing for universal quantum computation~\cite{nielsen_chuang}. Therefore, the quest is to reduce the error of both types of gates below the  fault-tolerance threshold in the presence of noise.

Among the different existing platforms for QIP, small crystals of trapped atomic ions in radio-frequency traps~\cite{review_qc_ions} are considered to be one of the most promising devices where these demanding thresholds could be achieved in the future. In fact, one-qubit gates with errors $\epsilon_{1{\rm q}}<\epsilon_{\rm FT_2}$ have already been demonstrated using trapped-ion hyperfine qubits subjected to microwave radiation~\cite{mw_ft_gates}. Additionally, two-qubit gates with errors as low as $\epsilon_{\rm FT_2}<\epsilon_{2{\rm q}}<\epsilon_{\rm FT_1}$ have also been demonstrated for trapped-ion optical qubits subjected to laser radiation~\cite{optical_ft_ms_gaes}. Since both qubits operate in a different frequency range (i.e. microwave/radio-frequencies versus optical), the reached accuracies have not been combined in a single device yet. Besides, it would be desirable to lower the error of the two-qubit gates even further to achieve $\epsilon_{2{\rm q}}<\epsilon_{\rm FT_2}$. 

In this work, we describe theoretically a scheme capable of achieving the two aforementioned goals. Our scheme is particularly designed for hyperfine trapped-ion qubits, enjoying thus the advantages of using microwave radiation for one-qubit gates. In addition, we introduce a scheme for a {\it driven geometric phase gate} acting on two qubits, which combines the advantages of microwave and laser radiation. This hybrid laser-microwave scheme can attain errors as low as $\epsilon_{2{\rm q}}<\epsilon_{\rm FT_2}$ in the presence of {\it (i)} thermal noise, {\it (ii)} dephasing noise, {\it (iii)} phase noise in the lasers, and {\it (iv)} intensity fluctuations of the microwave driving. In comparison to our previous proposal~\cite{ss_gate}, the protocol presented in this work allows us to increase the gate speed by at least one order of magnitude as a result of working closer to particular resonances of the laser-driven couplings. More importantly, such a {\it near-resonance regime}  turns out to be the parameter regime where the recent experimental demonstration~\cite{nist_ss_gate} of the two-qubit gate~\cite{ss_gate} has been performed~\cite{comment_exp}. Therefore, the present work will be useful to provide a theoretical background for the results presented in~\cite{nist_ss_gate}, where $\epsilon_{2{\rm q}}\approx\epsilon_{\rm FT_1}$ was achieved. Our analytical and numerical results show how the geometric character of these laser-microwave driven gates underlies the gate resilience to several sources of noise. Moreover, we optimize the laser-microwave parameters, such that the errors can be reduced below the fault-tolerance threshold $\epsilon_{2{\rm q}}<\epsilon_{\rm FT_2}$.

This article is organized as follows: In Sec.~\ref{driven_phase_gates}, we describe the mechanism for the driven geometric phase gates by means of detailed analytical analysis supported by numerical results. The relevant analytical calculations are presented in Appendix~\ref{app1}. In Sec.~\ref{noise_robustness}, the scheme is tested against several sources of noise (i.e. thermal, dephasing, phase, and intensity noise). Some of these noise sources can be modelled as a stochastic Hamiltonian term, whose properties are  described in Appendix~\ref{app2}.  Finally, we present some conclusions and prospects in Sec.~\ref{conc}.

\section{Driven single-sideband geometric phase gates}
\label{driven_phase_gates}

\begin{figure*}
\centering
\includegraphics[width=1.5\columnwidth]{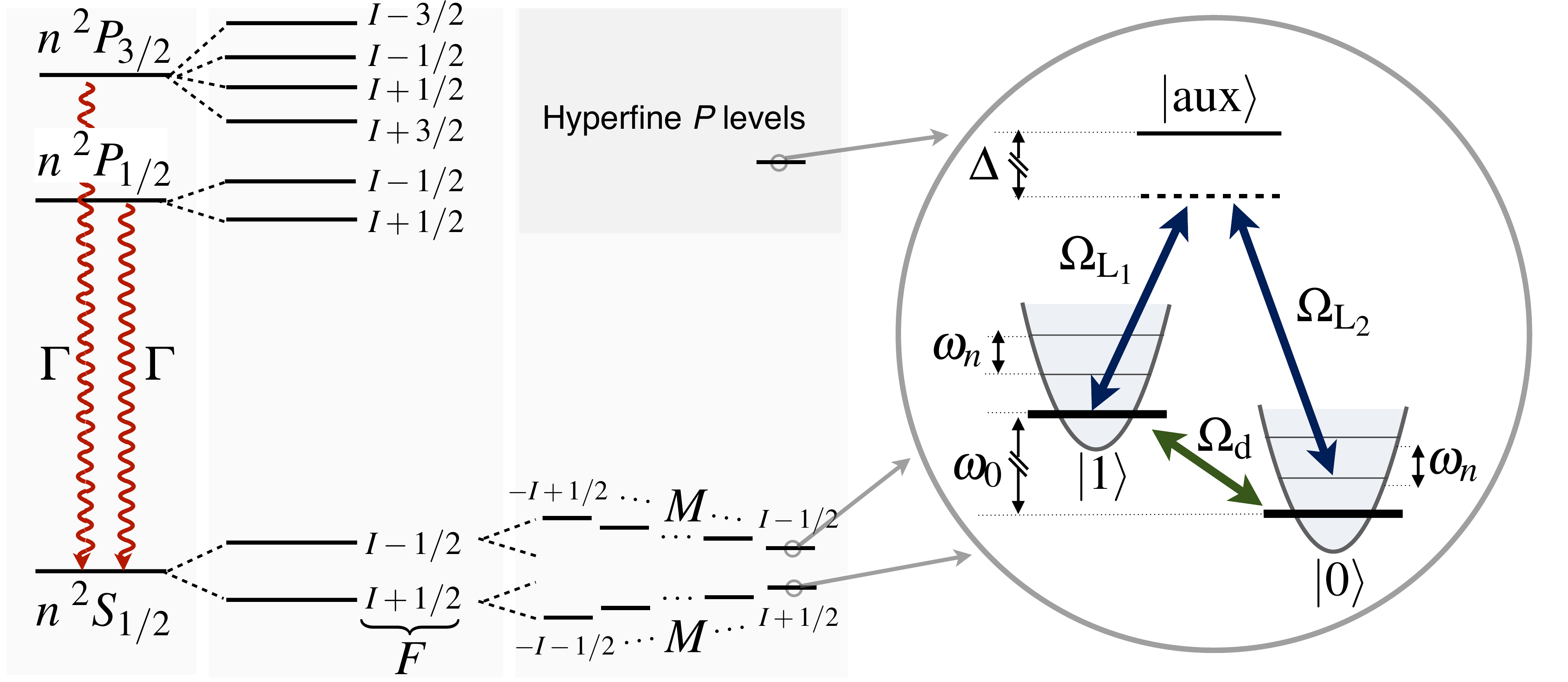}
\caption{{\bf Ingredients for the driven geometric phase gate:} For a generic ion species with hyperfine structure, as determined by the quantum numbers $n^2{\rm L}_JF$, the qubit is chosen from the two hyperfine states $F=I\pm\half$ of the ground-state manifold $n^2S_{1/2}$. Additionally, a quantizing magnetic field splits the $F$-states into Zeeman sub-levels, such that $\ket{0} \equiv \ket{F=I+\half,M }$ and $\ket{1} \equiv \ket{F'=I-\half, M'}$ define the hyperfine qubit.
Two laser beams with Rabi frequencies $\Omega_{\rm L_1},\:\Omega_{\rm L_2}$ drive the first red-sideband transition of the qubit in a stimulated Raman configuration $\ket{0}\otimes\ket{n}\to\ket{1}\otimes\ket{n-1}$, where $n$ is the number of phonons of a particular mode of the two-ion crystal. Simultaneously, the
qubit transition is driven directly by a microwave $\Omega_{\rm d}$.}
\label{fig_1}
\end{figure*}

\subsection{Two-ion crystals as the hardware for quantum logic gates}
\label{hardware}

Let us start by describing the system under consideration: a  two-ion ($N=2$) crystal confined in a linear Paul trap~\cite{wineland_review}. Under certain conditions~\cite{leibfried_review}, such radio-frequency traps provide an effective quadratic confining potential, which is characterized by the so-called axial $\omega_{z}$, and radial $\{\omega_x,\omega_y\}$ trap frequencies. Moreover, when
 $\{\omega_x,\omega_y\} \gg \omega_z$,  the ion equilibrium positions arrange in a string along the trap $z$-axis. As customary in these cases,  when the particles only perform small excursions around their equilibrium positions,  their motion can be described in terms of collective vibrational modes, whose quantized excitations lead to the well-known phonons~\cite{feynman_book}. In the  case of ions, the low-temperatures provided by laser-cooling techniques  justify such a treatment~\cite{james}, and the vibrations of the  crystal are described by 
 \begin{equation}
 \label{phonons}
 H_{\rm p}=\sum_{\alpha}\sum_{n=1}^{N}\omega_{\alpha,n}a_{n,\alpha}^{\dagger}a_{n,\alpha}^{\phantom{\dagger}},
 \end{equation}
 where  $a_{n,\alpha}^{\dagger}\:(a_{n,\alpha}^{\phantom{\dagger}})$ are the creation (annihilation) phonon operators, $\omega_{\alpha,n}$ is the vibrational frequency of the $n$-th normal mode along the axis $\alpha=\{x,y,z\}$, and we set $\hbar=1$ throughout this manuscript. According to Eq.~\eqref{phonons}, the vibrations along different axes are decoupled, which allows us  to focus our attention on the radial modes along the $x$-axis. However, we emphasize that our proposal also holds for the other modes.  To simplify notation, we write $a_{n,x}\to a_{n}$, $\omega_{n,x}\to \omega_{n}$. Note that for a two ion crystal there are two radial normal modes: the center-of-mass (com) and the zigzag (zz) mode which we also shall denote modes ''1`` and ''2``, respectively.
 
 In addition to the vibrational degrees of freedom, our two-ion crystal also  has internal (i.e. atomic) degrees of freedom. We consider ion species with a hyperfine structure, such that we can select two levels from the ground-state manifold to form our qubit $\{\ket{0_i},\ket{1_i}\}$. This particular choice of qubit has two important properties: {\it (i)} spontaneous emission between the qubit states is negligible (i.e. $T_1$ times are much larger than experimental time-scales), and {\it (ii)}  the typical qubit frequencies $\omega_0/2\pi=(E_1-E_0)/2\pi\sim$1-10\hspace{0.2ex}GHz allow for microwave, or radio-frequency, radiation to drive directly the qubit transition.
 To test our scheme with experimental realistic parameters, we shall consider $^{25}\text{Mg}^{+}$ in this work, although we remark that our scheme also works for other ion species (see Fig.~\ref{fig_1}). We choose two hyperfine states $\ket{0}\equiv \ket{F=3, M=3}$
 and $\ket{0}\equiv \ket{F'=2, M'=2}$, where $M,M'$ are the Zeeman sub-levels of the $F,F'$ levels of the electronic ground state manifold, which are separated by a frequency $\omega_0/2 \pi \approx 1.8\,$GHz.  
 
 By considering the standard Zeeman interaction between the atomic magnetic-dipole and an oscillating magnetic field from the microwave source, and using the standard rotating-wave approximation in quantum optics~\cite{scully}, the qubit Hamiltonian can be written as follows
 \begin{equation}
 \label{qubit}
 H_{\rm q}=\sum_{i=1}^N\half\omega_{0}\sigma_i^z+\half(\Omega_{\rm d}\sigma_i^+\ee^{-\ii\omega_{\rm d}t}+\text{H.c.}),
 \end{equation}
where we have introduced the qubit operators $\sigma_i^z=\ket{{1}_i}\bra{{1}_i}-\ket{{0}_i}\bra{{0}_i}$, $\sigma_i^+=\ket{{1}_i}\bra{{0}_i}= (\sigma_i^-)^{\dagger}$, and the frequency (Rabi frequency) $\omega_{\rm d}(\Omega_{\rm d})$ of the microwave driving. Note that the rotating wave approximation requires $\omega_{\rm d }\approx \omega_0$, and $|\Omega_{\rm d}|\ll\omega_0$, which still leaves the possibility of achieving very strong drivings $\Omega_{\rm d}/2\pi\sim$1-10\hspace{0.2ex}MHz, while benefiting from the intensity stability of microwave sources. We also note that the large wavelengths of the microwave traveling waves avoid any qubit-phonon coupling induced by the microwave, which would impose further constraints on the strengths of the  drivings. In this work, we exploit the possibility of obtaining such strong microwave drivings  to produce fast two-qubit geometric phase gates with robustness to several sources of noise.

 As done in previous two-qubit gate schemes~\cite{cz_gates,cz_gates_exp,ms_gates,ms_gates_nist,gphgates, jonathan_plenio_gates,didi_gate}, the main idea is to consider an additional coupling between  the qubits and  the phonons, and use the latter  to mediate interactions between distant qubits in the ion trap. We follow~\cite{ss_gate}, and consider  laser-induced qubit-phonon couplings~\cite{wineland_review}, which have already been demonstrated in several laboratories. Therefore, our work can be considered as  an instance of a hybrid laser-microwave protocol. For completeness, we remark that for specific qubit choices such qubit-phonon couplings can also be achieved with microwave radiation. It has been shown that in the near-field of microwave sources, it is possible to obtain qubit-phonon couplings by producing either static~\cite{static_gradients,static_gradients_exp}, or oscillating magnetic-field gradients~\cite{oscillating_gradients,oscillating_gradients_exp}. The generality of the scheme presented in the following subsection would allow us to use any of these approaches to achieve qubit-phonon coupling and thus, also to implement the scheme in an all-microwave protocol. Note, however, that we follow
none of these approaches~\cite{static_gradients,static_gradients_exp,oscillating_gradients,oscillating_gradients_exp} and therefore, the microwave radiation does not introduce any qubit-phonon coupling.

The qubit-phonon coupling is provided by a two-photon stimulated Raman transition via a third auxiliary level. 
The transitions to the auxiliary excited state are driven by two far detuned laser beams in a traveling-wave configuration, such that their detuning is much larger than the excited-state decay rate (e.g. for $^{25}\text{Mg}^{+}$ the linewidth of the excited state is $\Gamma/2\pi\approx 41.4$\hspace{0.2ex}MHz, and thus a detuning on the order of $\Delta/2\pi\approx10$-100\hspace{0.2ex}GHz would suffice, see Fig.~\ref{fig_1}). In this  limit, the excited state can be adiabatically eliminated~\cite{wineland_royal}, and the qubit-phonon coupling becomes
\begin{equation}
H_{\rm qp}=\sum \limits_{i} \half\Omega_{\rm L} \sigma_i^+ \ee^{\ii({\bf k}_{\rm L}\cdot{\bf r}_i- \omega_{\rm L} t)} + {\rm H.c.},
\label{raman}
\end{equation}
where $\Omega_{\rm L}=\Omega_{\rm L_1}\Omega_{\rm L_2}^*/2\Delta$ is the two-photon laser Rabi frequency, and the two-photon wavevector (frequency) ${\bf k}_{\rm L}={\bf k}_{\rm L_1}-{\bf k}_{\rm L_2}$ (${\omega}_{\rm L}={\omega}_{\rm L_1}-{\omega}_{\rm L_2}$) are defined in terms of the corresponding parameters of the two laser beams~\cite{note}. By directing ${\bf k}_{\rm L}$ along the $x$-axis, and setting the laser frequencies such that $\omega_{\rm L}\approx \omega_0 - \omega_n$, and the detunings $\delta_n= \omega_{\rm L} - (\omega_0 - \omega_n)$, such that $\delta_n \ll \omega_n$ and $|\Omega_{\rm L}|\ll\omega_n$, a rotating-wave approximation leads to the so-called first red-sideband excitation
\begin{equation}
H_{\rm qp}=\sum \limits_{i,n} \mathcal{F}_{in} \sigma_i^+ a_n \ee^{-\ii \omega_{\rm L} t} + {\rm H.c.}.
\label{qubit_phonon}
\end{equation}
Here, we have defined the red-sideband coupling strengths 
\begin{equation}
\label{forces}
\mathcal{F}_{i1}=\ii \frac{|\Omega_{\rm L}| \eta_1}{2\sqrt{2}},\hspace{1ex}\mathcal{F}_{i2}=\ii (-1)^i \frac{|\Omega_{\rm L}| \eta_2}{2\sqrt{2}},
\end{equation}
where we have set the Raman-beam phase $\varphi_{\rm L}=0$, and used the Lamb-Dicke parameters $\eta_n={\bf k}_{\rm L}\cdot{\bf e}_x/\sqrt{2m\omega_n}\ll1$. We will refer to these couplings generically as forces, since $\mathcal{F}_{in}\sqrt{2m\omega_n}$ corresponds to a force applied on the harmonic oscillator representing the vibrational mode.

As explained below, equations~\eqref{phonons},~\eqref{qubit} and~\eqref{qubit_phonon} form the {\it driven single-sideband Hamiltonian} 
\begin{equation}
\label{dss}
H_{\rm dss}=H_{\rm p}+H_{\rm q}+H_{\rm qp},
\end{equation}
 which can be used to obtain the desired geometric phase gates for strong-enough microwave drivings. At this point, it is worth commenting that strong qubit drivings have also been realized experimentally in combination with a state-dependent force~\cite{driving_ms}. The driving in this case endows the scheme with protection from dephasing noise, while the resilience to the thermal motion of the ions is provided by the state-dependent force. However, this scheme is prone to phase noise due to fluctuations in the laser-beam paths, or to intensity fluctuations of the qubit driving. In a different context, we should also mention that strong-driving-assisted protocols for the generation of entanglement have  also been considered theoretically for atoms in cavities~\cite{driving_cavities}, where the driving may  reduce errors due to thermal population of the cavity modes~\cite{thermal_cavities}.

\subsection{Driven geometric phase gates}
\label{magnus}

\subsubsection{Introduction: geometric phase gates with a single sideband}

In reference~\cite{ss_gate}, we have analyzed how the driven single-sideband Hamiltonian $H_{\rm dss}$ leads to an {\it entangling gate} capable of producing  two Bell states at particular instants of time
\begin{equation}
\label{table}
\begin{split}
\textstyle{
\ket{01}\to\ket{\Psi^-}=\frac{1}{\sqrt{2}}(\ket{01}-\ii\ket{10}),}\\
\textstyle{\ket{10}\to\ket{\Psi^+}=\frac{1}{\sqrt{2}}(\ket{01}+\ii\ket{10}),}\\
\end{split}
\end{equation}
  in the far-detuned regime $|\mathcal{F}_{in}| \ll \delta_n$.  In this limit, the weak qubit-phonon coupling~\eqref{qubit_phonon} is responsible for second-order processes where phonons are virtually created and annihilated. This term alone leads to a flip-flop qubit-qubit   interaction $H_{\rm int}\propto J^{\rm{eff}}_{ij} \sigma_i^+\sigma_j^-+\text{H.c.}$  via virtual phonon exchange, where the coupling strength is given by 
	$J^{\rm{eff}}_{ij} = - \sum_n \mathcal{F}_{in} \mathcal{F}_{jn}^*/\delta_n$ with the above introduced forces and detunings $\mathcal{F}_{in}$ and $\delta_n$, respectively.	At certain instants of time  this interaction can act as an entangling gate. However, we also showed there that the performance of such a gate would be severely limited by thermal and dephasing noise.  The main idea put forward in~\cite{ss_gate} was to exploit a strong resonant microwave driving~\eqref{qubit}, fulfilling $\omega_{\rm d}=\omega_0$ and $\delta_n\ll\Omega_{\rm d}\in\mathbb{R}$, as  a continuous version~\cite{cdd,cdd_nv} of refocusing spin-echo sequences~\cite{pulsed_dd,pulsed_dd_ions}, to protect the two-qubit gate from these sources of noise. In this case, the virtual phonon exchange leads to an Ising-type interaction $H_{\rm int}\propto \tilde{J}^{\rm{eff}}_{ij} \sigma_i^x\sigma_j^x+\text{H.c.}$, where we have introduced $\sigma_i^x=\sigma_i^++\sigma_i^-$ and $\tilde{J}^{\rm{eff}}_{ij} = 1/4\,J^{\rm{eff}}_{ij}$. This qubit-qubit interaction can generate  all of the four Bell states
\begin{equation}
\label{table_bis}
\begin{split}
\textstyle{\ket{00}\to\ket{\Phi^-}=\frac{1}{\sqrt{2}}(\ket{00}-\ii\sgn(\tilde{J}^{\rm{eff}}_{12})\ket{11}),}\\
\textstyle{\ket{01}\to\ket{\Psi^-}=\frac{1}{\sqrt{2}}(\ket{01}-\ii\sgn(\tilde{J}^{\rm{eff}}_{12})\ket{10}),}\\
\textstyle{\ket{10}\to\ket{\Psi^+}=\frac{1}{\sqrt{2}}(\ket{01}+\ii\sgn(\tilde{J}^{\rm{eff}}_{12})\ket{10}),}\\
\textstyle{\ket{11}\to\ket{\Phi^+}=\frac{1}{\sqrt{2}}(\ket{00}+\ii\sgn(\tilde{J}^{\rm{eff}}_{12})\ket{11}).}\\
\end{split}
\end{equation}
Unfortunately, the explored far-detuned regime leads to gates that  are more than one order of magnitude slower than state-of-the-art implementations based on other schemes~\cite{gate_review}. Therefore, although  both the simplicity of the gate, and its  resilience to different sources noise are interesting advantages, its lower speed presents a considerable drawback. 

In this work, we show how one can abandon the far-detuned regime, while still preserving the nice properties of the driven single-sideband gate. Below, we show that by working in the context of the geometric phase gates~\cite{ms_gates, gphgates,ozeri_review}, our  {\it driven nearly-resonant single-sideband gate}: {\it (i)} Can attain speeds which are one order of magnitude faster than the far-detuned gate~\cite{ss_gate} for comparable parameters. For the specific parameters considered in this manuscript
(see Table~\ref{tab_1}) we can attain a gate speed of $t_{\rm g}\sim63\,\mu$s. Gate speeds in the range $10$-$100\,\mu$s can  be expected for for other parameters, or ion species.  {\it (ii)} Can minimize thermal and dephasing errors down to $\epsilon_{\rm th},\epsilon_{\rm d} <10^{-4}$, which improves by more than one order of magnitude the far-detuned gate~\cite{ss_gate}. {\it (iii)} Can withstand fluctuations of the laser phase occurring on timescales longer than the 63$\,\mu$s,  such that $\epsilon_{\rm ph} <10^{-4}$, which was outlined in~\cite{ss_gate}, but not analyzed carefully. {\it (iv)} Can also resist relative fluctuation in the intensity of the microwave driving at the $10^{-4}$-level directly with errors $\epsilon_{\rm I}\sim 10^{-3}$. Moreover, we show that by adding a secondary driving, the scheme can support stronger intensity fluctuations, while providing smaller gate errors $\epsilon_{\rm I}< 10^{-4}$.

\subsubsection{Qualitative analysis: dressed-state interaction picture and rotating-wave approximation}

To understand the mechanism underlying the {\it driven nearly-resonant single-sideband gate}, let  us make a small detour, and consider the so-called geometric phase gates by state-dependent forces~\cite{ms_gates, gphgates,ozeri_review}. By combining the red-sideband ($\sigma_i^+a_n$) term~\eqref{qubit_phonon} with a blue-sideband ($\sigma_i^+a_n^{\dagger}$) that has an opposite detuning but equal strength and adjusting the laser phases appropriately~\cite{ms_gates}, the qubit-phonon Hamiltonian becomes
\begin{equation}
\label{state_dep_force}
\hat{H}_{\rm qp}=\sum_{i,n}\mathcal{F}_{in}\sigma_i^xa_n\ee^{-\ii \delta_nt}+\text{H.c.},
\end{equation}
where the "hat" refers to the interaction picture with respect $H_0=H_{\rm q}+H_{\rm p}$, and we have switched off the microwave driving $\Omega_{\rm d}=0$. This hamiltonian~\eqref{state_dep_force} can be understood as a pushing force  in a direction that depends on the qubit state in the $x$-basis $\sigma_i^x\ket{\pm_i}_x=\pm\ket{\pm_i}_x$. Since a single Pauli matrix appears in the above equation, the time evolution under such a state-dependent pushing force reduces to that of a forced quantum harmonic oscillator~\cite{book_merz}, which can be solved exactly (see e.g.~\cite{transverse_phonons_phase_gate}). This leads to the following time evolution operator
\begin{equation}
\label{time_evolution}
\hat{U}(t)=\ee^{\sum_{in}\left(\frac{\mathcal{F}_{in}}{\delta_n}(\ee^{-\ii\delta_nt}-1)\sigma_i^xa_n-\text{H.c.}\right)}\ee^{+\ii\sum_{ijn}\frac{\mathcal{F}_{in}\mathcal{F}^*_{jn}}{\delta_n}(t-\frac{\sin(\delta_nt)}{\delta_n})\sigma_i^x\sigma_j^x}.
\end{equation}
The first unitary corresponds to a displacement operator, which  is well-known in quantum optics~\cite{scully}, with the peculiarity of being state-dependent. In phase space, this term induces periodic circular trajectories for the vibrational modes that depend on the collective spin state, and thus leads to qubit-phonon entanglement. When $t_{\rm g}=k_n 2\pi/\delta_n,\:k_n \in \mathbb{Z}$, the trajectories close and the qubits and phonons become disentangled, such that the evolution operator becomes 
\begin{equation}
\hat{U}(t_{\rm g})=\ee^{-\ii t_{\rm g}\sum_{ij}J_{ij}^{\rm MS}\sigma_i^x\sigma_j^x},\hspace{2ex} J_{ij}^{\rm MS}=\textstyle{-\sum_n\frac{\mathcal{F}_{in}\mathcal{F}^*_{jn}}{\delta_n}}.
\end{equation}
This unitary can be easily seen to provide the desired entangled states~\eqref{table_bis} when $t_g (2J_{12}^{\rm eff})=\pi/4$. Additionally, if the initial spin states are eigenstates of $\sigma_1^x\sigma_2^x$, the gate gives the table
\begin{equation}
\label{table_phase_gate}
\begin{split}
&\textstyle{\ket{++}_x\to\phantom{\ee^{\ii\frac{\pi}{2}}}\ket{++}_x,}\\
&\textstyle{\ket{+-}_x\to\ee^{\ii\frac{\pi}{2}}\ket{+-}_x,}\\
&\textstyle{\ket{-+}_x\to\ee^{\ii\frac{\pi}{2}}\ket{-+}_x,}\\
&\textstyle{\ket{--}_x\to\phantom{\ee^{\ii\frac{\pi}{2}}}\ket{--}_x,}\\
\end{split}
\end{equation}
up to an irrelevant global phase. This corresponds to a two-qubit $\pi/2$-phase gate which, together with single-qubit rotations, gives a universal set of gates for quantum computation. The fact that the phase-space trajectory is closed at $t_{\rm g}$, allows for the interpretation of the $\pi/2$-phases as geometric Berry phases determined by the area enclosed by the trajectory~\cite{ozeri_review}. 
Let us remark two properties of these gates: {\it (i)} They do not rely on any far-detuned condition $|\mathcal{F}_{in}|\ll\delta_n$, and can be thus much faster, {\it (ii)} The geometric origin of the phase gate  underlies its robustness with respect to thermal motion of the ions (i.e. spin-phonon disentanglement occurs when the trajectory closes regardless of the vibrational state).

After this small detour, it becomes easier to understand why a gate based only on a nearly-resonant single-sideband cannot lead to a geometric phase gate, and is thus very sensitive to thermal noise. By rewriting the single-sideband~\eqref{qubit_phonon} in the interaction picture with respect to $H_0$ for $\Omega_{\rm d}=0$, we get
\begin{equation}
\label{single_sideband_ip}
\hat{H}_{\rm qp}=\sum \limits_{i,n} \half\mathcal{F}_{in} (\sigma_i^x+\ii\sigma_i^y) a_n \ee^{-\ii \delta_n t} + {\rm H.c.},
\end{equation}
where $\sigma_i^y=-\ii\sigma_i^++\ii\sigma_i^-$. Hence, it is clear that the red-sideband yields a combination of two  state-dependent forces acting on orthogonal bases. This fact forbids an exact solution, such as the one obtained for a single  state-dependent  force~\eqref{time_evolution}, and we inevitably lose  the notion of state-dependent trajectories that close independently of the vibrational state. Accordingly, the qubit and phonons get more entangled as the far-detuned condition $|\mathcal{F}_{in}|\ll\delta_n$ is abandoned, and the gate fidelity drops severely for ions in thermal motion.

If we now switch on a resonant microwave driving $\omega_{\rm d}=\omega_0,\Omega_{\rm d}\in\mathbb{R}$, then the  Hamiltonian~\eqref{single_sideband_ip} changes into
\begin{equation}
\label{single_sideband_ip2}
\tilde{H}_{\rm qp}\!=\!\sum \limits_{i,n}\! \frac{\mathcal{F}_{in}}{2} (\!\sigma_i^x+\ii\sigma_i^y\cos(\Omega_{\rm d}t)-\ii\sigma_i^z\sin(\Omega_{\rm d}t)\!) a_n \ee^{-\ii \delta_n t} + {\rm H.c.}.
\end{equation}
Here, the "tilde" refers to the following "dressed-state" interaction picture, namely $\tilde{H}_{\rm qp}=\tilde{U}_{\rm d}(t){H}_{\rm qp}\tilde{U}^{\dagger}_{\rm d}(t)$, where  
\begin{equation}
\label{dressed_basis}
\tilde{U}_{\rm d}(t)=\ee^{\ii t\sum_i\half\Omega_{\rm d}\sigma_i^x}\ee^{\ii t \sum_i\half\omega_0\sigma_i^z}\ee^{\ii t\sum_n\omega_na_n^{\dagger}a_n}.
\end{equation} 
It is  now possible to argue that the state-dependent $\sigma^y$ and $\sigma^z$ forces rotate very fast for sufficiently strong drivings $\Omega_{\rm d}\gg\delta_n$, and can be thus neglected in a rotating wave approximation provided that $|\mathcal{F}_{in}|\ll\Omega_{\rm d}$. Under this constraint, the driven single-sideband Hamiltonian becomes
\begin{equation}
\label{x_force}
\tilde{H}_{\rm qp}\approx\sum \limits_{i,n}\half \mathcal{F}_{in} \sigma_i^x a_n \ee^{-\ii \delta_n t} + {\rm H.c.},
\end{equation}
and thus a formal solution like Eq.~\eqref{time_evolution} becomes possible again, provided that we make the substitution $\mathcal{F}_{in}\to \half \mathcal{F}_{in}$. 

According to this qualitative argument, a strongly-driven single-sideband Hamiltonian can also produce the desired geometric phase gate. Therefore, it seems possible to abandon the far-detuned regime $|\mathcal{F}_{in}|\ll\delta_n$ of our previous work~\cite{ss_gate}, and obtain faster gates that are still robust with respect to thermal noise. In the following sections, we will present a quantitative detailed analysis to test the validity of this idea.

\subsubsection{ Quantitative  analysis:  Magnus expansion and numerical analysis}

As discussed above, the presence of different state-dependent forces in  the Hamiltonian~\eqref{single_sideband_ip} avoids an exact solution for the unitary time evolution operator. However, since we are interested in the strong-driving limit $ |\mathcal{F}_{in}|\ll \Omega_{\rm d} $,  we can use the so-called Magnus expansion~\cite{me_ref}, truncating it to the desired order  to obtain the leading contributions to the dynamics (see Appendix~\ref{app1} for the details). The  time-evolution operator in the Schr\"{o}dinger picture reads 
\begin{equation}
\label{magnus_u}
U_{\rm app}(t)=\tilde{U}^{\dagger}_{\rm d}(t)\ee^{\Omega(t)},\hspace{1.5ex}\Omega(t)\approx{\Omega_1(t)+\Omega_2(t)+\mathcal{O}(\xi,\chi)},
\end{equation}
 and $\xi=(\Omega_{\rm L}\eta_{n})^2/\Omega_{\rm d}^2$, $\chi=(\Omega_{\rm L}\eta_{n})^2/\Omega_{\rm d}\delta_n$ are the small parameters in the Magnus expansion. In this expression,  we have defined the  anti-unitary operators containing the different state-dependent displacements
\begin{equation}
\begin{split}
  \Omega_1(t)&=  \sum \limits_{i,n} \frac{\mathcal{F}_{in}}{2\delta_n} ( \ee^{-\ii \delta_n t} -1)  \sigma_i^x a_n-\text{H.c.}   \\
  &+ \sum \limits_{i,n} \frac{\mathcal{F}_{in}}{4(\Omega_{\rm d}-\delta_n)} ( \ee^{\ii (\Omega_{\rm d} - \delta_n) t} -1)  (-\ii\sigma_i^y+\sigma_i^z) a_n -\text{H.c.}  \\
    &+ \sum \limits_{i,n} \frac{\mathcal{F}_{in}}{4(\Omega_{\rm d}+\delta_n)} ( \ee^{-\ii (\Omega_{\rm d} + \delta_n) t} -1)  (\ii\sigma_i^y+\sigma_i^z) a_n -\text{H.c.}, \\
	\end{split}
	\label{Omega1}
	\end{equation}
	and the operators leading to the qubit-qubit interactions
	\begin{equation}
	\begin{split}
  \Omega_2(t)=& \ii\sum_{i,j,n}{\frac{\mathcal{F}_{in}\mathcal{F}_{jn}^*}{4\delta_n}}\sigma_i^x \sigma_j^x\bigg(t-\frac{\sin\delta_nt}{\delta_n}\bigg)\\
   +&\ii t \sum_{i,n}\!\Delta\Omega_{in}\!\left(\!\!a_n^{\dagger}a_n-\frac{1}{2}\!\right)\sigma_i^x+\sum_{i,n\neq m}(f_{nm}(t)\sigma_i^xa_m^{\dagger}a_n-{\rm H.c.}),
  \end{split}
	\label{Omega2}
	\end{equation}
where we have  introduced the couplings strengths
\begin{equation}
\label{shifts}
	\Delta\Omega_{in} = -\frac{|\mathcal{F}_{in}|^2}{4} \left( \frac{1}{\Omega_{\rm d} - \delta_n} + \frac{1}{\Omega_{\rm d} + \delta_n}  \right),
\end{equation}
and the following time-dependent functions
\begin{equation}
\label{residual_spin_phonon}
	f_{nm}(t) = \frac{\mathcal{F}_{jn}\mathcal{F}_{jm}^*}{8(\delta_n - \delta_m)}\!\! \left( \frac{1}{\Omega_{\rm d} - \delta_m} + \frac{1}{\Omega_{\rm d} + \delta_m}\right)\!\!\big(\ee^{-\ii(\delta_n-\delta_m)t} - 1 \big).
	\end{equation}

At this point, it is worth comparing the final expressions~\eqref{Omega1}-\eqref{Omega2} to those corresponding to a state-dependent force in Eq.~\eqref{time_evolution}. The first-order contribution,~\eqref{Omega1}, resembles  the state-dependent displacement in~\eqref{time_evolution}, but we get in addition the contribution of more state-dependent forces in different bases.  From the first line of the  second-order contribution~\eqref{Omega2}, we observe also a similarity with the qubit-qubit couplings in~\eqref{time_evolution}, but we get additional residual qubit-phonon couplings in the second line of~\eqref{Omega2}. We  discuss below how all these additional terms can be minimized, such that we are left with an effective time evolution that is analogous to that of a state-dependent force.

\begin{figure}
\centering
\includegraphics[width=0.75\columnwidth]{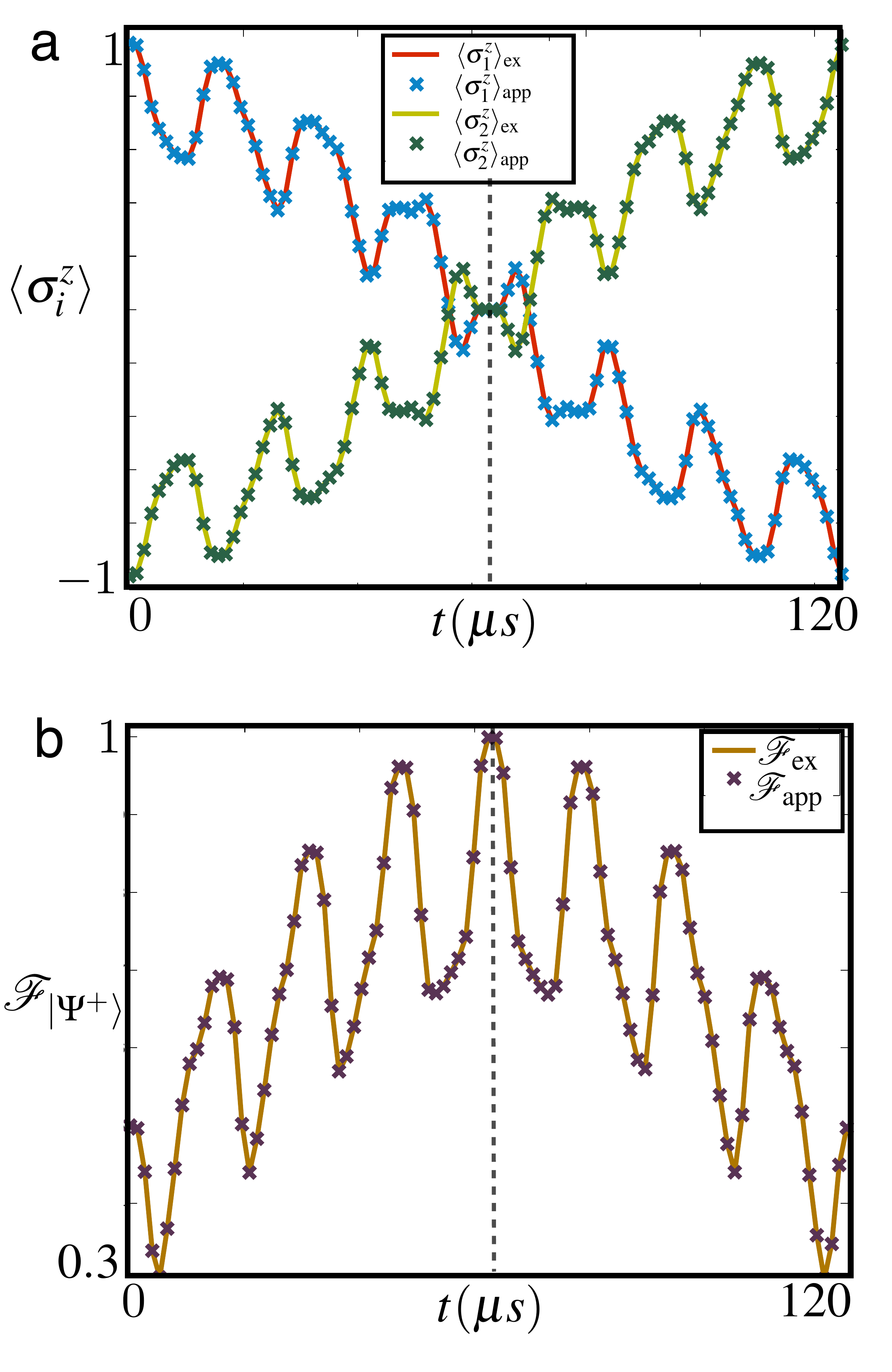}
\caption{{\bf Entangling $\sigma_1^x\sigma_2^x$ gate:} The figure displays the exact dynamics by  numerical integration of the unitary evolution~\eqref{exact_u} (solid lines) with the dynamics given by the approximated time-evolution operator (crosses). We consider an initial state $\rho_0=\ketbra{10}\otimes\rho_{\rm th}(\bar{n}_1,\bar{n}_2)$, where $\bar{n}_1\approx\bar{n}_2\approx 0$ are the initial mean number of phonons for the cm and zz modes. Note that due to the nearly-resonant sideband, phonons are created during the gate, and we have to set a high truncation $n_{\rm max}=10$ to the vibrational Hilbert spaces.   In {\bf (a)}, we represent the dynamics of local qubit operators $\langle \sigma_i^z\rangle$, which show the qubit flip $\ket{10}\to\ket{01}$ after $t=126\,\mu$s.  In {\bf (b)}, we display the fidelity between the time-evolved state  and the Bell state $\ket{\Psi^+}=(\ket{01}+\ii\ket{10})/\sqrt{2}$. At $t=63\,\mu$s, this fidelity approaches unity. The agreement of both descriptions supports the validity of our analytical derivation.}
\label{fig_magnus_qubit}
\end{figure}

Let us note that Eqs.~\eqref{Omega1} and~\eqref{Omega2} correspond to  the leading terms in the second-order expansion expansion of $U(t)$, while the complete expression may be found in  Appendix~\ref{app1}. Before analyzing how the geometric phase gates arise from Eqs.~\eqref{Omega1}-\eqref{Omega2}, let us check numerically the validity of our derivation  by comparing it to the time-evolution under the full driven single-sideband Hamiltonian $H_{\rm dss}$~\eqref{dss}. To perform the numerics more efficiently, we  expressed $H_{\rm dss}$ in a picture  where it becomes time independent, namely
\begin{equation}
\label{dss_time_indep}
H_{\rm dss}'=\sum_n\delta_na_n^{\dagger}a_n+\sum_i\frac{\Omega_{\rm d}}{2}\sigma_i^x+\sum_{i,n}(\mathcal{F}_{in}\sigma_i^+a_n+\text{H.c.}),
\end{equation}
where $H_{\rm dss}'=U'(t)H_{\rm dss}(U'(t))^{\dagger}$, and the unitary is
\begin{equation}
{U'}(t)=\ee^{\ii t\sum_n(\omega_0-\omega_{\rm L})a_n^{\dagger}a_{n}}\ee^{\ii t \sum_i\half\omega_0\sigma_i^z}.
\end{equation} 
Accordingly, the time-evolution operator can be written  as
\begin{equation}
\label{exact_u}
U_{\rm exact}(t)=({U'}(t))^{\dagger}\ee^{-\ii H_{\rm dss}' t}.
\end{equation}
 For the numerical simulations, we chose realistic parameters for ion-trap experiments, which  are summarized in Table~\ref{tab_1}. The outcome of the simulations is shown in Fig.~\ref{fig_magnus_qubit}{\bf (a)}, which shows a very good agreement between the Magnus expansion and the exact time evolution  for the qubit dynamics. This supports the validity of our derivations, and allows us to carry on with the description of the driven geometric phase gate.

\begin{figure}
\centering
\includegraphics[width=0.75\columnwidth]{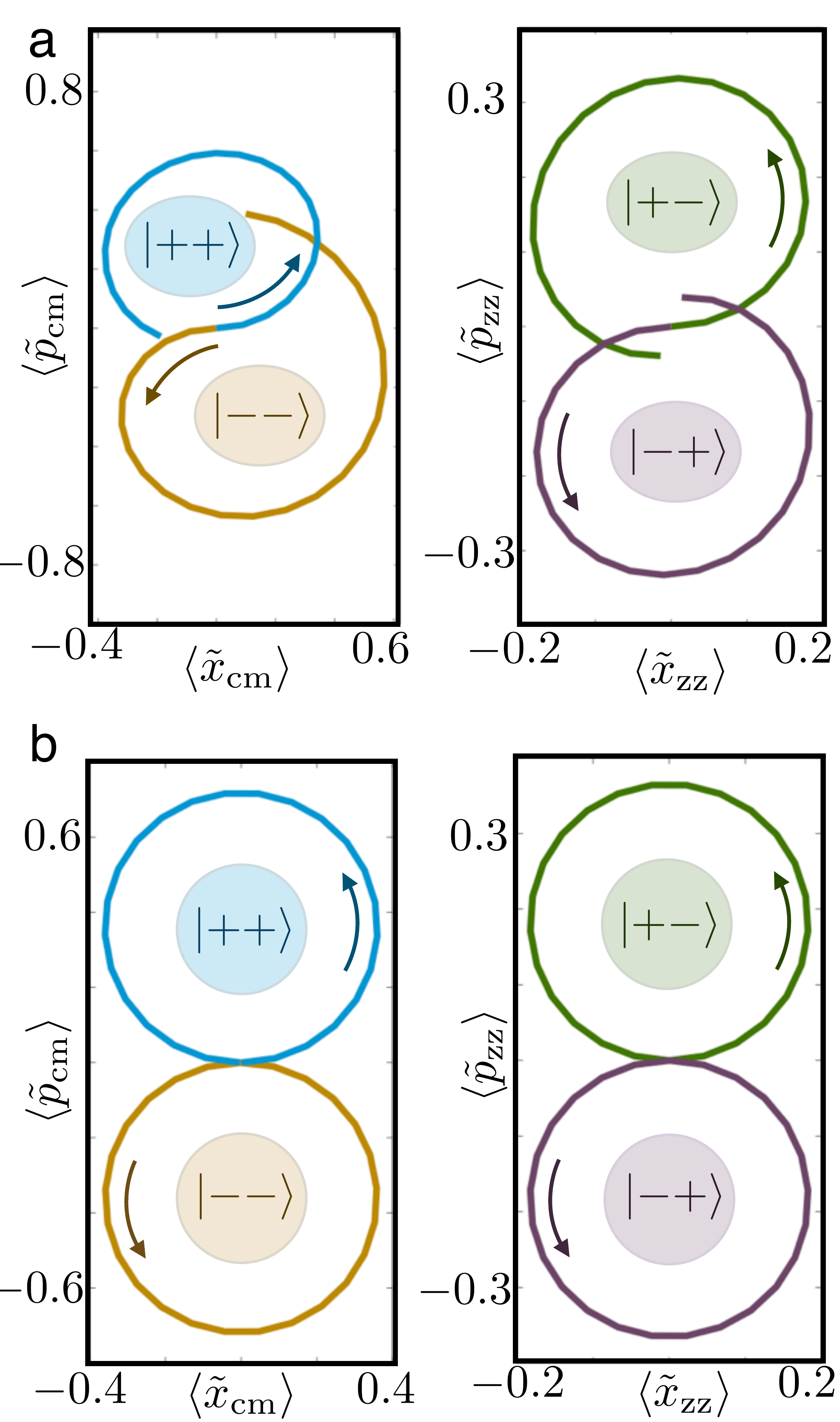}
\caption{{\bf Phase-space trajectories:} Numerical analysis of the phase-space trajectories  under the full driven single-sideband Hamiltonian~\eqref{dss}. {\bf (a)} Trajectories of the forced normal modes in the weak driving regime $\Omega_{\rm d}/2\pi=200\,$kHz. In the left panel, the initial spin state is $\ket{++}$ or $\ket{--}$ while the phonons initially are in a thermal state $\rho_{\rm th}(\bar{n}_1,\bar{n}_2)=\rho_{\rm th}(\bar{n}_1)\otimes\rho_{\rm th}(\bar{n}_2)$ with $\bar{n}_2 \approx \bar{n}_1=0.5$, and according to the forces~\eqref{forces}, only the center-of-mass mode is driven in phase space $\langle\tilde{x}_{\rm cm}\rangle=1/\sqrt{2}\langle a_1+a_1^{\dagger}\rangle,\langle\tilde{p}_{\rm cm}\rangle=\ii/\sqrt{2}\langle a_1^{\dagger}- a_1\rangle$. In the right panel, the initial spin state is $\ket{+-}$ or $\ket{-+}$, such that only the zigzag mode develops a driven trajectory in this case $\langle\tilde{x}_{\rm zz}\rangle=1/\sqrt{2}\langle a_2+a_2^{\dagger}\rangle,\langle\tilde{p}_{\rm zz}\rangle=\ii/\sqrt{2}\langle a_2^{\dagger}- a_2\rangle$. In both cases, the trajectories are not closed. {\bf (b)} Same as {\bf (a)} but in the strong-driving limit $\Omega_{\rm d}/2\pi=5\,$MHz. As an effect of the strong driving, the trajectories are closed, and the gate inherits a geometric character. In the numerics we have to set a truncation $n_{\rm max} = 15$ to the vibrational Hilbert spaces.}
\label{fig_trajectories}
\end{figure}

The action of the first contribution $\Omega_1(t)$~\eqref{Omega1} to the
 time-evolution operator~\eqref{magnus_u} can be understood as follows: for each vibrational mode, three non-commuting state-dependent forces aim at displacing the ions along circular paths in phase space, such that the specific direction of each trajectory depends on the particular eigenstates $\ket{\pm_x},\:\ket{\pm_y}$ and $\ket{0/1}$ of the operators $\sigma^x,\:\sigma^y$ and $\sigma^z$. For weak drivings, the combination of these forces will deform the circular paths, and lead to phase space trajectories that are not closed any longer (see Fig.~\ref{fig_trajectories}{\bf (a)}). 
 It is clear from the analytical expression~\eqref{Omega1} that, by applying a sufficiently-strong driving $\Omega_{\rm d}\gg\delta_n,\:|\mathcal{F}_{jn}|$, the forces in the $\sigma^{y/z}$-basis get suppressed, and, to a good approximation, we are left with the desired single state-dependent force in the $\sigma^x$-basis with a halved strength $\mathcal{F}_{in}\to \half \mathcal{F}_{in}$ with respect to the standard M\o lmer-S\o rensen term~\eqref{state_dep_force}. Accordingly, we should recover the circular phase-space trajectories when the initial state of the qubits is an eigenstate of the $\sigma_1^x\sigma_2^x$ operator. In Fig.~\ref{fig_trajectories}{\bf (b)},  we analyze the strong-driving dynamics numerically, and confirm the above prediction. Therefore, the qualitative description of the previous section is put on a firmer ground by the use of the Magnus expansion. 
 
\begin{table}
\centering
  \caption{{\bf Values of trapped-ion setup for the numerical simulation}  }
\begin{tabular}{ c  c c  c  c  c  c  c}
\hline \hline
  $\omega_z/2 \pi$ & $\omega_x/2 \pi$ & $\delta_{{\rm com}}/2\pi$ &$\eta_{{\rm com}}$ & $\delta_{{\rm zz}}/2\pi$ & $\eta_{{\rm zz}}$& $\Omega_{\rm L}/2 \pi$& $\Omega_{\rm d}/2 \pi$ \\
\hline
$1\,$MHz \hspace{0.1ex} & $4\,$MHz  \hspace{0.1ex} &  $127\,$kHz & \hspace{0.1ex} 0.225 & $254\,$kHz&0.229 & $811\,$kHz  & 7.2 {\rm MHz}  \\
\hline 
\hline
\end{tabular}
\label{tab_1}
\end{table}

The Magnus expansion also allows us to improve further the geometric character of the gate. By setting the correct values to the detunings and intensities, it is possible to cancel exactly the qubit-phonon entanglement introduced by all three state-dependent displacements in~\eqref{Omega1}, and the residual qubit-phonon entanglement introduced in the second term of the second line of Eq.~\eqref{Omega2}. In order to achieve this, we need to adjust the Hamiltonian parameters such that,  at the gate time $t_{\rm g}$, the following constraints are fulfilled
\begin{equation}
\label{constraints}
\begin{split}
t_{\rm g}= r \frac{2 \pi}{\delta_1}, \hspace{1ex} &\delta_2=k\delta_1, \hspace{1ex}\Omega_{\rm d}=p\delta_1, \hspace{2ex}\: r,k,p\in \mathbb{Z}, \\
&\frac{r}{k} \in \mathbb{Z} \hspace{1ex} {\rm and} \hspace{1ex} |p|> |r|+|k|.
\end{split}
\end{equation}
In other words, we need the detuning of the zigzag mode $\delta_{2}$ and the microwave Rabi frequency $\Omega_{\rm d}$ to be an integer multiple of the center-of-mass mode detuning $\delta_1$. 
Physically, the constraints can be understood in the following way. By fulfilling Eq.~\eqref{constraints} the center-of-mass mode performs $r$ closed loops in phase space during one gate cycle while the zigzag mode performs $r/k$ loops in phase space.
Thus, in order to attain the geometric character of the gate $r/k$ must be an integer for then the phase space trajectories of both normal modes close simultaneously at $t_{\rm g}$. The constraints on $p$, however, lack a clear physical interpretation. They are motivated by the results of the Magnus expansion.% If the constraints on $p$ are met, almost all contributions of the Magnus expansion vanish exactly at the gate time.

For the above conditions, $\Omega_1(t_{\rm g})=0$ vanishes exactly. Moreover, we reduce considerably the qubit-phonon entanglement due to the second-order term $  \Omega_2(t_{\rm g})= -\ii t_{\rm g} H_{\rm dss}$, where we have introduced the Hamiltonian
\begin{equation}
\label{eff_H_qubit_phonon}
 H_{\rm dss}=\sum_{i,j}J_{ij}^{\rm dss}\sigma_i^x \sigma_j^x-\sum_{i,n}\Delta\Omega_{in}(a_n^{\dagger}a_n-\half)\sigma_i^x,
\end{equation}
and the following coupling strengths
\begin{equation}
\label{def_couplings}
J_{ij}^{\rm dss}=-\sum_n \frac{1}{4\delta_n}\mathcal{F}_{in} \mathcal{F}_{jn}^{*}.
\end{equation}
 Note that a closer inspection of all the additional terms in the Magnus expansion (see Appendix~\ref{app1}) shows that they all cancel exactly under these constraints. Therefore, all the dynamics for the gate is exactly described by the Hamiltonian~\eqref{eff_H_qubit_phonon}.

Finally, in the limit of very strong drivings $\Omega_{\rm d}\gg\delta_n$, the residual qubit-phonon coupling is minimized as $\Delta\Omega_{in}$~\eqref{shifts} decreases with increasing driving. 
Moreover, one can even cancel exactly the contribution of the second term in Eq.~\eqref{eff_H_qubit_phonon} by introducing a spin-echo pulse~\cite{ss_gate} or, equivalently, a phase reversal~\cite{nist_ss_gate} of the microwave driving
at half the expected gate time. As an additional benefit, this procedure guarantees the closure of phase space trajectories in the case of an imperfect detuning if the forces in the $\sigma^{y/z}$ bases
 can be neglected in a rotating wave approximation. This has been demonstrated in the recent experiment~\cite{nist_ss_gate}.
Therefore, the full time evolution can be written approximately as
\begin{equation}
\label{eff}
U_{\rm app}(t_{\rm g})=\tilde{U}^{\dagger}_{\rm d}(t_{\rm g})\ee^{-\ii H_{\rm dss}t_{\rm g}},\hspace{2ex}
 H_{\rm dss}=\sum_{i,j}J^{\rm{dss}}_{ij}\sigma_i^x\sigma_j^x.
\end{equation}
 From this expression, we can extract the gate time to be $(2J_{12}^{\rm dss})t_{\rm g}=\pi/4$, which will give rise to the entangling table~\eqref{table_bis}, or to the phase-gate table~\eqref{table_phase_gate}, depending on the chosen basis for the initial states. In Fig.~\ref{fig_magnus_qubit}{\bf (b)}, we check the validity of this description by computing the fidelity of generating the Bell state $\ket{\Psi^+}$ at the expected gate time, starting from the initial qubit state $\ket{10}$. As seen from the figure, the agreement between the full Magnus expansion and the exact dynamics is very good, and we obtain the desired entangled state at $t_{\rm g}$ with very high fidelities. It can also be observed in the figure that the fidelity exhibits 
oscillations. These can be understood from Eq.~\eqref{time_evolution} which we realize to a good approximation with our choice of the driving strength. The state $\ket{10}$ couples to the state-dependent forces on both modes and each of the modes is
displaced in phase space. In this process entanglement between the internal and motional states is created and the fidelity of producing the Bell state drops. Whenever a trajectory is closed the entanglement between qubit and motional states vanishes and
the fidelity exhibits a local maximum. The oscillations observed in the fidelity can then be understood as the beatnote of the two state dependent forces, each acting on a different mode.

Note that we have not made any assumption on the ratio between the sideband strengths and the laser detunings $|\mathcal{F}_{in}|/\delta_n$, and thus we do not require to work in the far-detuned regime $|\mathcal{F}_{in}|/\delta_n\ll1$ of our previous proposal~\cite{ss_gate}. According to the expression of the gate coupling strengths $J^{\rm{dss}}_{ij}$, this means that we can increase the gate speed considerably by working closer to the  resonance (i.e. $t_{\rm g}\approx 63\,\mu$s). The conditions~\eqref{constraints} also imply that the phase space trajectories of the vibrational modes are closed, providing a geometric character to the two-qubit  gate, and its resilience to the thermal noise.

 In the following sections, we will analyze   the  speed and noise robustness  of this {\it driven nearly-resonant single-sideband gate}. However, let us first  address how the constraints~\eqref{constraints} can be met in practice. The detunings for the center-of-mass and the zigzag modes are given by
\begin{equation}
\label{eq:par2.3}
\delta_{1} =  \omega_{\rm L} - (\omega_0 -\omega_{1} ),\hspace{2ex}
\delta_{2} =  \omega_{\rm L} - (\omega_0 -\omega_{2} ).
\end{equation}
By setting $\delta_1 = c \omega_1$, such that $c$ is a constant, and demanding that $\delta_2 = k \delta_1$ with $k\in\mathbb{Z}$,  we get
\begin{equation}
c = \frac{\xi -1 }{k -1},
\label{eq:par2.4}
\end{equation}
where $\xi=\omega_2/\omega_1<1$ is the ratio of the normal-mode frequencies. To make the gate as fast as possible $t_{\rm g}=2\pi/(|c|\omega_1)$, the modulus of $c$ should be 
as big as possible, and we thus choose $k=2$, so that $|c|=1-\xi$. According to these considerations, the gate time in units of the trap frequency is $t_{\rm g}\omega_x=2\pi/(1-\xi)$, which is minimized by minimizing $\xi$ (i.e. large difference of the center-of-mass and zigzag mode). This occurs when approaching the linear-to-zigzag instability.

 Finally, the laser intensities in  $J^{\rm{dss}}_{12}$ have to be adjusted such that we can fulfil the following condition
\begin{equation}
t_{\rm g}=\frac{\pi}{8 J^{\rm{dss}}_{12}} = r \frac{2 \pi}{\delta_1},
\label{eq:par2.5}
\end{equation}
where $r\in \mathbb{Z}$ is an integer that determines how many loops in phase space the center-of-mass mode performs. On the one hand, $r$ should be as small as possible to maximize the gate speed. On the other hand, it turns out that the smaller $r$ is, the more fragile the gate becomes with respect to dephasing noise (see sections below). Therefore, we optimize the value of $r$ to find a compromise between gate speed and gate fidelity.

Inserting  expression~\eqref{def_couplings} in the above equation, we obtain
\begin{equation}
|\Omega_{\rm L}|=\frac{|\delta_1|}{\eta_1 \sqrt{\frac{r}{2} \left(1-\frac{1}{2 \xi} \right)}},
\label{eq:par2.6}
\end{equation}
which fixes the laser intensities. Finally, we note that  the microwave driving strength can be modified in the experiment
\begin{equation}
\Omega_{\rm d}=p\delta_1,
\end{equation}
 with $p\in\mathbb{Z}$ being a very large integer to meet the strong-driving condition $\Omega_{\rm d}\gg\delta_n$. With these expressions, the parameters of the setup are fixed in terms of two integers $(r,p)$. In the following section, we optimize the choice of these integers to maximize simultaneously the gate fidelity and speed. We explore how such fidelities approach the fault-tolerance threshold in the presence of different sources of  noise  as the microwave driving strength is increased.

\section{Gate robustness against different sources of noise}
\label{noise_robustness}

In this section, we provide a detailed study of the behavior of the gate in the presence of noise. We will analyze four different sources of noise: {\it (i)} {\it Thermal noise}, which is caused by the thermal motion of the ions (i.e. we do not assume perfect ground-state cooling of the radial modes). {\it (ii)} {\it Dephasing noise}, which is caused by fluctuating Zeeman shifts due to non-shielded magnetic fields, or by fluctuating ac-Stark shifts due to non-compensated energy shifts caused by fluctuating laser intensities. {\it(iii)} {\it Phase noise}, which is caused by fluctuations in the phases of the laser beams. {\it (iv)} {\it Intensity noise}, which is caused by fluctuations in the microwave driving strength. We will show that the driven single-sideband gate is intrinsically robust to noise of the type {\it (i), (ii), (iii)}, whereas it will suffer from {\it (iv)} at very strong drivings. We propose  to incorporate an additional weaker driving to make the gate also robust with respect to intensity fluctuations.
To show all these properties, we integrate numerically the dynamics~\eqref{exact_u} given by the full Hamiltonian~\eqref{dss_time_indep} subjected to the additional noise sources. In the simulations, the parameter $r$ 
in~\eqref{eq:par2.5} and~\eqref{eq:par2.6} is set to $r=8$, since we have found that it gives the best compromise between fidelity and speed.

\subsection{Resilience to thermal noise}

One of the primary sources of gate infidelities in early trapped-ion proposals~\cite{cz_gates} was the noise introduced by residual thermal motion of the ions after a stage of resolved-sideband cooling. In order to consider such thermal noise, we  consider an initial state $\rho_0=\ketbra{10} \otimes \rho_{\rm th}(\bar{n}_1,\bar{n}_2)$ for a thermal phonon state $\rho_{\rm th}(\bar{n}_1,\bar{n}_2)=\rho_{\rm th}(\bar{n}_1)\otimes\rho_{\rm th}(\bar{n}_2)$ characterized by the mean phonon numbers $\{\bar{n}_1,\bar{n}_2\}$ for each mode. Note that due to the small difference in frequency of the normal modes in x-direction we have $\bar{n}_1 \approx \bar{n}_2$. According to the table~\eqref{table_bis}, the target state after the gate is $\rho(t_{\rm g})=\ketbra{\Psi^+} \otimes \rho_{\rm th}(\bar{n}_1,\bar{n}_2)$, where $\ket{\Psi^+}=(\ket{01}+\ii\ket{10})/\sqrt{2}$.

Based on the Magnus-expansion evolution operator~\eqref{magnus_u}, it is possible to derive a closed expression for the gate fidelity
\begin{equation}
\mathcal{F}_{\ket{\Psi^+}}=\tr\{ \ket{\Psi^+} \bra{\Psi^+} \otimes \mathbbm{1}_{{\rm ph}} U_{\rm M}(t_{\rm g})\rho_0 U_{\rm M}^{\dagger}(t_{\rm g})\}.
\end{equation}
 in the strong driving limit $\Omega_{\rm d}\gg \delta_n$ (i.e. neglecting all terms that are suppressed with $1/\Omega_{\rm d}$). After some algebra, the fidelity can be expressed as a sum of expectation values of the displacement operator
 \begin{equation}
 \langle D(\alpha_n)\rangle={\rm Tr}_{\rm ph}\{\ee^{\alpha_n a_n^{\dagger}-\alpha_n^* a_n}\rho_{\rm th}(\bar{n}_n)\}=\ee^{-|\alpha_n|^2(\bar{n}_n+\half)}.
 \end{equation}
Provided that the gate time fulfills $t_{\rm g}\Omega_{\rm d}=4\pi p$ where $p\in\mathbb{Z}$, the fidelity of the entangling operation reads as follows
\begin{equation}
\mathcal{F}_{\ket{\Psi^+}}=\fourth+\half\ee^{-\sum_n\kappa_n(1-\cos(\delta_n t_{\rm g}))}+\frac{1}{8}\sum_n\ee^{-4 \kappa_n(1-\cos(\delta_n t_{\rm g}))}
\end{equation}
where $\kappa_n=\frac{|\mathcal{F}_{1n}|^2}{\delta_n^2}(2\bar{n}_n+1)$. From this expression, it becomes clear that the fidelity would decrease exponentially with the mean number of thermal phonons, unless the phase-space trajectories are closed $\delta_n t_{\rm g}=2\pi$. Under such condition, the gate fidelity is maximized $\mathcal{F}_{\ket{\Psi^+}}(t_{\rm g}=2\pi/\delta_n)=1$ in the limit of $\Omega_{\rm d}\to \infty$. This expression not only unveils the importance of the geometric character of the gate, but will also be useful in understanding the effect of other sources of noise.

 In order to check how the fidelity approaches unity as the microwave driving $\Omega_{\rm d}$ is increased, we study numerically the exact time evolution~\eqref{exact_u}. To avoid timing
errors due to the fast oscillations induced by the strong microwave driving, we add { the above mentioned} refocusing spin-echo pulse at half the expected gate time, such that the total time evolution is  $U_{\rm full} (t_{\rm g},0)= U_{\rm exact}(t_{\rm g},\frac{t_{\rm g}}{2})(\sigma^z_1 \sigma_2^z) U_{\rm exact}(\frac{t_{\rm g}}{2},0)$.
We then compute numerically the fidelity of producing the desired Bell state 
\begin{equation}
\mathcal{F}_{\ket{\Psi^+}} = \tr\{ \ket{\Psi^+} \bra{\Psi^+} \otimes \mathbbm{1}_{{\rm ph}} U_{\rm full}(t_{\rm g},0) \rho_0 U_{\rm full}^{\dagger} (t_{\rm g},0)\}.
\end{equation} 
 In order to integrate the full Hamiltonian numerically, we truncate each vibrational Hilbert space to a maximum number of phonons of $n_{\rm max}=25$ per mode, which is sufficiently high so that no appreciable error is introduced.

\begin{figure}
\centering
\includegraphics[width=0.9\columnwidth]{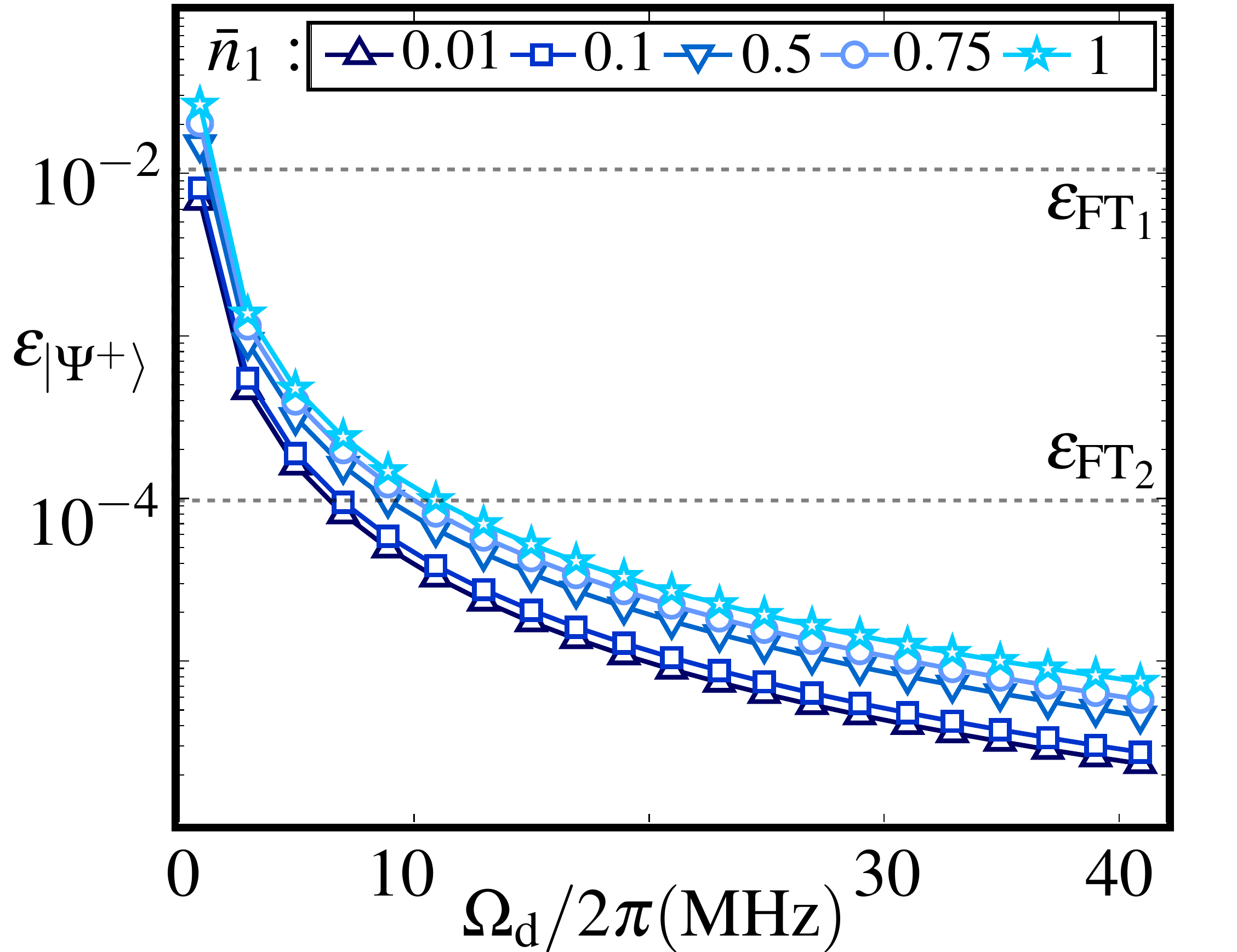}
\caption{{\bf Thermal noise:} Error $\epsilon = 1 - \mathcal{F}$ of producing the Bell state $\ket{\Psi^+}$ from the initial state $\ket{10}$ at the expected gate time $t_{\rm g}$ as a function of the applied
microwave driving field $\Omega_{\rm d}$. Note that each point  of the figures fulfills the condition $\Omega_{\rm d}=p\delta_1$ for an integer $p$. The thermal phonon population of the initial state  $\bar{n}_1$ is increased from $0$ to $1.$ Gate errors  well-below the fault-tolerance threshold (FT) can be achieved $\epsilon_{2{\rm q}}<\epsilon_{\rm FT_2}$. }
\label{fig_thermal}
\end{figure}

In Fig.~\ref{fig_thermal}, we represent the numerical results for the gate error $\epsilon=1-\mathcal{F}$ for various microwave driving strengths. By  setting $r=8$ in Eq.~\eqref{eq:par2.5}, we get  a gate time of $t_g \approx 63 \,\mu$s, which improves the gate speed of our previous work~\cite{ss_gate} by one order of magnitude.
Besides, the results shown in figure \ref{fig_thermal} show that the gate errors surpass the fault-tolerance threshold $\epsilon_{{\rm FT}_2}\sim 10^{-4}$ for sufficiently strong drivings. This figure  illustrates clearly the benefit of this nearly-resonant single-sideband gate in comparison to our previous scheme~\cite{ss_gate}, since the gate error even for the largest thermal phonon numbers $\bar{n}_1=1$ is reduced by several orders of magnitude.

\subsection{Resilience to dephasing noise}
\label{dephasing_section}

Another source of gate infidelities for several trapped-ion qubits is the noise caused by fluctuating magnetic fields, or laser intensities, 
which lead to a shift in the resonance frequency of the qubit through either the Zeeman shift, or an uncompensated ac-Stark shift (see Appendix~\ref{app2}). 
 On the time scales considered, these fluctuations can be modelled as 
\begin{equation}
H_{\rm fluc} (t)=\sum_i \half \Delta \omega_0(t) \sigma_i^z,
\label{eq:par3.1}
\end{equation}
where $\Delta \omega_0(t) $ is a normal stochastic Markov process known as an Ornstein-Uhlenbeck (O-U) process, and we have assumed that the fluctuating source acts globally on the two ions (e.g. global fluctuating magnetic field). 

The O-U process, which is defined by a relaxation time $\tau$ and a diffusion constant $c$~\cite{ouproc}, is characterized by the following identities
\begin{equation}
\overline{\Delta\omega_0(t)}=0,\hspace{2ex}\overline{\Delta\omega_0(t)\Delta\omega_0(t-s)}=\textstyle{\frac{c\tau}{2}}\ee^{-\frac{s}{\tau}},
\end{equation}
where the "bar" refers to the stochastic average, and we assume a Markovian limit where the timescales of interest greatly exceed the correlation time. The time-evolution of the density matrix in the so-called Born-Markov approximation 
\begin{equation}
\label{BM}
\dot{\rho}=-\int_0^{\infty}{\rm d}s\overline{[H_{\rm fluc}(t),[H_{\rm fluc}(t-s),\rho(t)]]},
\end{equation}
can be expressed in the following Lindblad form
\begin{equation}
\label{dephasing}
\dot{\rho}=\mathcal{L}_{\rm d}({\rho})=\Gamma_{\rm d}(S_z\rho S_z-\half S_z^2\rho-\half\rho S_z^2),\hspace{2ex}\Gamma_{\rm d}={\frac{c\tau^2}{4}}
\end{equation}
where we have introduced the   collective operator $S_{\alpha}=\sum_i\sigma_i^{\alpha}$ for $\alpha=z$. By means of the adjoint master equation~\cite{breuer_book}, one finds that $\sigma_i^x(t)=\ee^{\mathcal{L}^{\dagger}_{\rm d}t}\sigma_i^x(0)=\sigma_i^x(0)\ee^{-t c\tau^2/2}$, 
which allows us to define the decoherence time as $T_2 = 2/c \tau^2$. This expression, together with the condition of short noise correlation times,  fulfilled by setting $\tau=0.1T_2$, allows us  to set the noise-model parameters  for the study of a variety of decoherence times $T_2\in[15,40]\mu {\rm s}$. We note that these values are overly pessimist, and would only occur for a poor shielding of the magnetic fields, or a bad stabilization of the lasers.

 As we show below, even for these poor coherences, the strong microwave driving suppresses the noise, and reestablishes a good coherent behavior. The qualitative argument is again to move to the dressed-state interaction picture~\eqref{dressed_basis}, where $\tilde{H}_{\rm fluc} (t)=\tilde{U}_{\rm d}(t){H}_{\rm fluc} (t)\tilde{U}_{\rm d}^{\dagger}(t)$ becomes
\begin{equation}
\tilde{H}_{\rm fluc} (t)= \sum \limits_{i} \half\Delta \omega_0(t) \left( \cos(\Omega_{\rm d} t) \sigma_i^z + \sin(\Omega_{\rm d} t) \sigma_i^y \right).
\label{eq:par3.2}
\end{equation}
For a sufficiently strong driving, the noisy terms rotate very fast, and  we can neglect them in a rotating-wave approximation.

In order to put this argument on a firmer footing, let us use again the Born-Markov approximation~\eqref{BM}. We obtain a Lindlad-type master equation $\dot{\tilde{\rho}}=\tilde{\mathcal{L}}_{\rm d}(\tilde{\rho})$, where the new dephasing super-operator is
\begin{equation}
\label{decoupled_deph}
\tilde{\mathcal{L}}_{\rm d}(\tilde{\rho})=-\ii[\frac{\Delta\Omega_{\rm d}}{2}S_x,\tilde{\rho}]+\frac{\tilde{\Gamma}_{\rm d}}{2}\!\!\sum_{\alpha=z,y}\!\!(S_{\alpha}\rho S_{\alpha}-\half S_{\alpha}^2\rho-\half\rho S_{\alpha}^2),
\end{equation}
and we have assumed that $\Omega_{\rm d}T_2\gg1$. Here, we have defined a dressed-state energy shift and a renormalized dephasing rate
\begin{equation}
\Delta\Omega_{\rm d}=\frac{\Omega_{\rm d}\tau}{4T_2(1+(\Omega_{\rm d}\tau)^2)},\hspace{2ex}\tilde{\Gamma}_{\rm d}=\frac{\Gamma_{\rm d}}{(1+(\Omega_{\rm d}\tau)^2)}.
\end{equation}
The adjoint master equation associated to such Liouvillian~\eqref{decoupled_deph}  yields the coherences  decay $\sigma_i^x(t)=\sigma_i^x(0)\ee^{-2\tilde{\Gamma}_{\rm d}t}$. Accordingly, we get a renormalized decoherence time
\begin{equation}
\tilde{T}_2=T_2(1+(\Omega_{\rm d}\tau)^2),
\end{equation}
which  increases quadratically with the driving strength~\cite{diss_gates}.

\begin{figure}
\centering
\includegraphics[width=0.9\columnwidth]{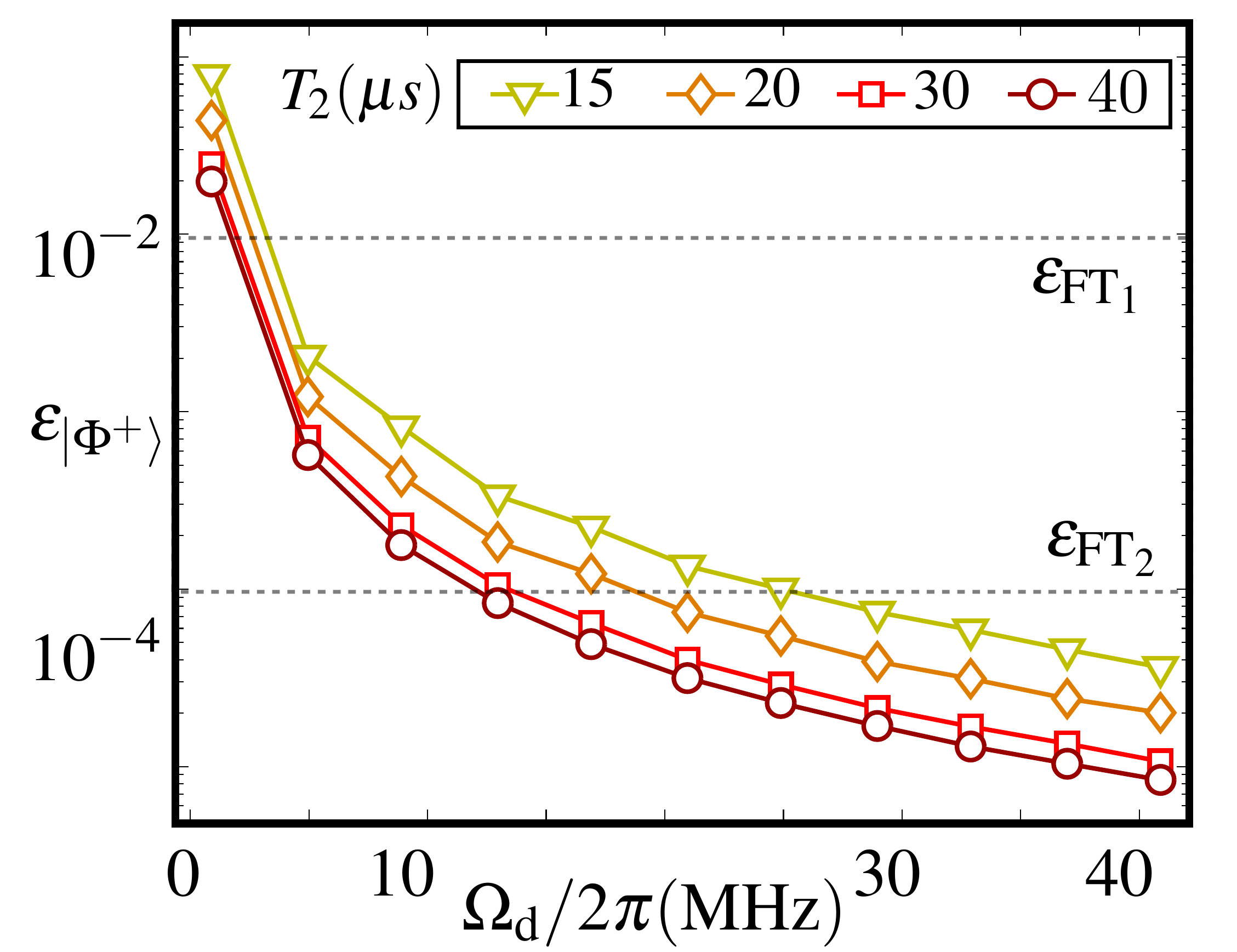}
\caption{ {\bf Dephasing noise:} Error in the generation of the Bell state $\ket{\Phi^-}$ from the initial state $\ket{00}\otimes\ket{\bar{n}_1,\bar{n}_2}$ with $\bar{n}_n=0$, for a time evolution including the dephasing noise. The results shown include a statistical average over $N_{\rm n}=10^3$ realizations of the noise process per point. We set a truncation of the vibrational Hilbert spaces to hold a maximum of $n_{\rm max}=7$ phonons per mode. The error is shown as a function of the applied microwave driving strength, and for different values of $T_2 \in [15\mu {\rm s}, 40 \mu {\rm s}]$. Gate errors $\epsilon_{2{\rm q}}$ below the fault-tolerance threshold (FT) can be achieved $\epsilon_{2{\rm q}}<\epsilon_{\rm FT_2}$.  }
\label{fig_dephasing_noise}
\end{figure}

We now address qualitatively the noise effects on the fidelity of the entangling operation in the strong-driving limit. Note that both $\ket{10}$ and  $\ket{\Psi^+}$ lie in the "zero-magnetization" subspace, which can be easily checked to be  a decoherence-free subspace of the original dephasing~\eqref{dephasing}. Therefore, the effects of the dephasing noise must be tested for another Bell state, such as 
$\ket{\Phi^-}=(\ket{00}- \ii\ \ket{11})/\sqrt{2}$, which is generated from $\ket{00}$ (see Eq.~\eqref{table_bis}). We identify two possible effects: {\it (i)} On the one hand,  the qubit coherences are degraded in a timescale given by $\tilde{T}_2=T_2(1+(\Omega_{\rm d}\tau)^2)$. {\it (ii)} On the other hand, the state-dependent forces for the leading term in~\eqref{magnus_u} are damped according to $\mathcal{F}_{in}\to\mathcal{F}_{in}\ee^{-t/\tilde{T}_2}$ (i.e. recall $\sigma_i^x(t)=\sigma_i^x (0)\ee^{-2\tilde{\Gamma}_{\rm d}t}$). This avoids the perfect closure of the phase-space trajectories, and thus decreases the gate fidelity.

To validate these predictions,  we integrate the full Hamiltonian~\eqref{dss_time_indep}  incorporating the noise
term~\eqref{eq:par3.1} { and once again a refocusing spin-echo pulse at half the gate time}. In Appendix~\ref{app2}, we describe in detail the conditions under which the stochastic noise process can be numerically propagated in time, and incorporated into the full time evolution. In Fig.~\ref{fig_dephasing_noise}, we show the numerical results for the gate error as a function of the microwave driving strength, where we have set $\bar{n}_1=\bar{n}_2=0$ to distinguish clearly between the effects of thermal and dephasing noise.
This figure  shows neatly how   the dephasing  error is considerably suppressed by increasing the driving strength. Errors well-below
the fault-tolerance threshold can be achieved again for sufficiently-strong drivings. Additionally, this figure also shows the advantage of 
the near-resonant gates with respect to the far-detuned ones~\cite{ss_gate}, since the error  is reduced  by several orders of magnitude.

%For the fastest possible (setting $r=2$, see eqs. \eqref{eq:par2.5} and \eqref{eq:par2.6}) the gate error still lies below the faul tolerance threshold but doesn't 
%reach $10^{-3}$. The is because due to the magnetic noise the $\mathcal{F}_{jn}$ giving the coupling strengths and consequently the gate speed will start to fluctuate and trajectories cannot be closed exactly. 
%The error is caused by the contribution of $\propto \mathcal{F}_{jn}/\delta_n$ of \eqref{eq:par2.1} which grows with an increasing Rabi frequency which is needed for faster gates (see \eqref{eq:par2.6}).

\subsection{Resilience to phase noise}

Another possible source of noise for trapped-ion gates are fluctuations in the laser phases at the position of the ions. Such fluctuations become especially dangerous for the geometric phase gates based on two non co-propagating Raman laser beams, which are more sensitive to laser path fluctuations. This noise can be modelled by substituting the sideband couplings $\mathcal{F}_{in}\to\mathcal{F}_{in}\ee^{\ii\Delta\varphi(t)}$, where $\Delta\varphi(t)$ is again a stochastic variable. Therefore, the full Hamiltonian~\eqref{dss_time_indep} becomes
\begin{equation}
\label{dss_phase_noise}
H_{\rm dss}'=\sum_n\delta_na_n^{\dagger}a_n+\sum_i\frac{\Omega_{\rm d}}{2}\sigma_i^x+\sum_{i,n}(\mathcal{F}_{in}\ee^{\ii\Delta\varphi(t)}\sigma_i^+a_n+\text{H.c.}).
\end{equation}
However, in this case we model the noise as another normal stochastic process, namely a Wiener process $\Delta\varphi\in[0,2\pi]$. This stochastic process is fully characterized by a single diffusion constant $c$ and its
initial value
\begin{equation}
\label{def_wiener_process}
\overline{\Delta\varphi(t)}=\Delta\varphi(0)=0,\hspace{2ex}\overline{\Delta\varphi(t)\Delta\varphi(t)}=c t.
\end{equation}
As phase fluctuations typically occur on a much longer time-scale as compared to the typical gate times~\cite {exp_phase_noise_schmidt_kaler}, they contribute with slow phase drifts. Therefore, we simulated $N_{\rm n}=10^3$ subsequent realizations of the gate where the phase drifts were
modelled as a Wiener process with 
\begin{equation}
c=\frac{(\zeta_{\rm p}\pi)^2}{10^3 t_g},\hspace{2ex}\zeta_{\rm p}\in[0,0.1]
\end{equation}
where the parameter $\zeta_{\rm p}$ determines how large the phase drifts are. A more detailed discussion of the process, and the motivation for the choice of the diffusion constant, can be found in Appendix~\ref{app2}.

Let us note that our analytical study~\eqref{magnus_u} predicts that the gate is insensitive to such slow phase drifts. In this regime, we can set the phase to be constant $\Delta\varphi(t)=\Delta\varphi_0$  during the gate interval. Accordingly, we can use the analytic expression in Eqs.~\eqref{Omega1} and~\eqref{Omega2} after substituting $\mathcal{F}_{in}\to\mathcal{F}_{in}\ee^{\ii\Delta\varphi_0}$, where $\Delta\varphi_0\in[0,2\pi]$ is a time-independent random variable. From these expressions, one can see that: {\it (i)} The geometric character of the gate, and thus its robustness to thermal noise, is not altered by any constant value of $\Delta\varphi_0$ (i.e. the fulfilment of the constraints~\eqref{constraints}, such that $\Omega_1(t_{\rm g})=0$, is independent of $\Delta\varphi_0$). {\it (ii)} The effective qubit-qubit interactions~\eqref{eff}, which determine the gate time, $J_{ij}^{\rm dss}=\sum_{n}(\mathcal{F}_{in}\ee^{\ii\Delta\varphi_0})(\mathcal{F}_{jn}\ee^{\ii\Delta\varphi_0})^*/4\delta_n=\sum_{n}\mathcal{F}_{in}\mathcal{F}_{jn}^*/4\delta_n$ are independent of the phase value. Therefore, we conclude that for strong-enough drivings and sufficiently slow phase drifts, the gate should be robust against phase noise. 

\begin{figure}
\centering
\includegraphics[width=0.9\columnwidth]{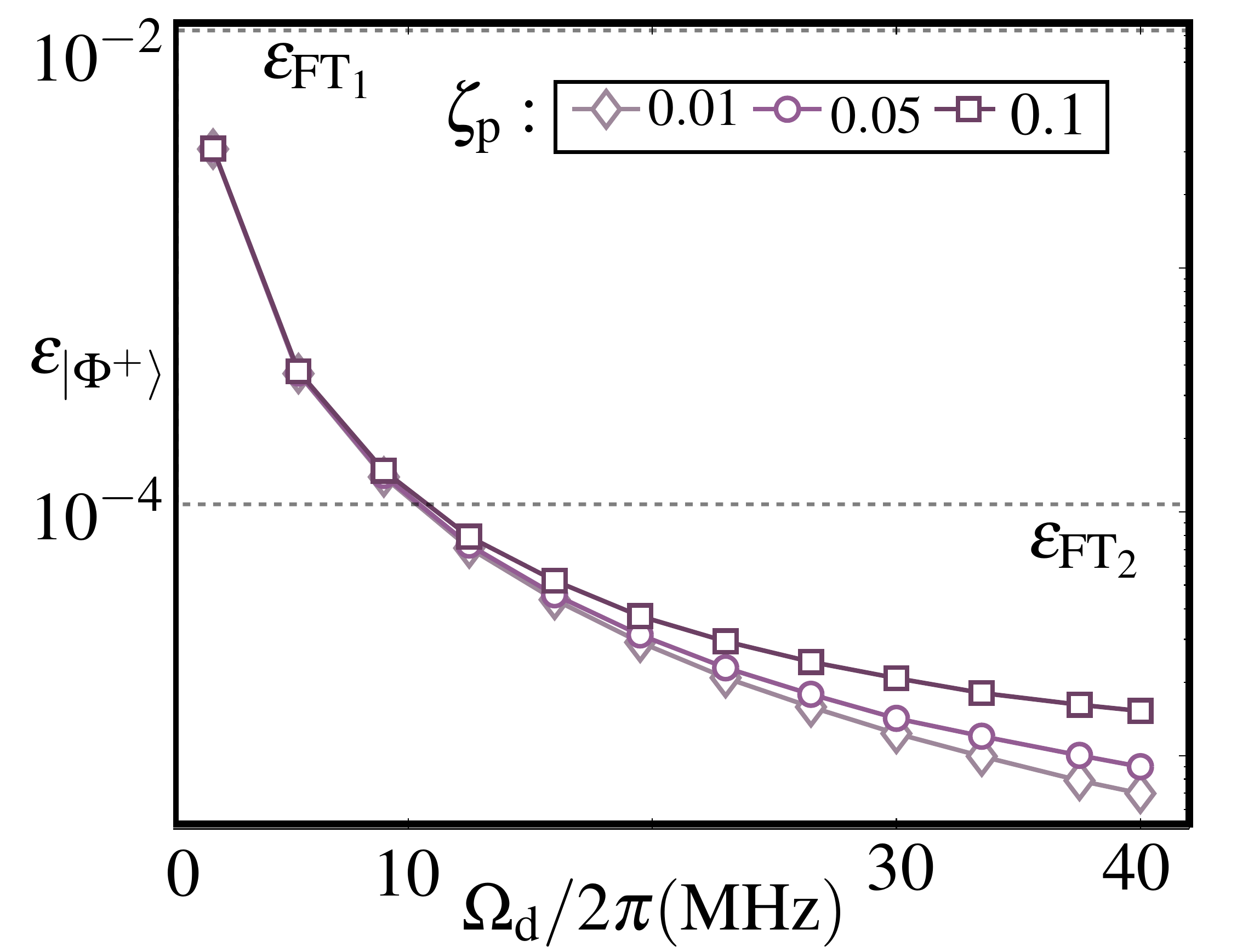}
\caption{ {\bf Phase noise:} Error in the generation of the Bell state $\ket{\Phi^-}$ from the initial state $\ket{00}\otimes\ket{\bar{n}_1,\bar{n}_2}$ with $\bar{n}_n=0$, for a time evolution including slow noisy drifts in the laser phases. The results shown include a statistical average over $N_{\rm n}=10^3$ subsequent realizations of the gate in the presence of the noise process~\eqref{def_wiener_process}, and we set a truncation of the vibrational Hilbert spaces to hold a maximum of $n_{\rm max}=7$ phonons per mode. The error is shown as a function of the applied microwave driving strength, and for different values of the variance of the noise associated to the parameter $\zeta_{\rm p}$. Gate errors  below the fault-tolerance threshold (FT) can be achieved $\epsilon_{2{\rm q}}<\epsilon_{\rm FT_2}$.  }
\label{fig_phase_noise}
\end{figure}

In Fig.~\ref{fig_phase_noise}, we present our numerical results for the resilience of the gate to slow drifts in the phases of the lasers. From these results, we can conclude that for moderate drivings, and phases with drifts as big as $\Delta\varphi\approx 0.1\pi$ over $10^3$ gate realizations, the gate errors still lie well-below the stringent fault-tolerance threshold $\epsilon_{2{\rm q}}<\epsilon_{\rm FT_2}$. We thus confirm that the proposed gate is indeed robust against a realistic phase noise, where the phase drifts for the time-scales of interest are well below $\Delta\varphi\approx 0.1\pi$ (see the discussion in Appendix~\ref{app2}). 

\subsection{Partial resilience to intensity noise}

Let us now discuss the impact of possible  fluctuations of the microwave intensity on the gate fidelities. This noise is modelled by substituting the microwave Rabi frequency $\Omega_{\rm d}\to\Omega_{\rm d}(t)=\Omega_{\rm d}+\Delta\Omega_{\rm d}(t)$, where $\Delta\Omega_{\rm d}(t)$ is a stochastic process representing the microwave intensity fluctuations. Hence, the full Hamiltonian~\eqref{dss_time_indep} should be substituted for
\begin{equation}
\label{dss_intensity_noise}
H_{\rm dss}'=\sum_n\delta_na_n^{\dagger}a_n+\sum_i\frac{\Omega_{\rm d}(t)}{2}\sigma_i^x+\sum_{in}(\mathcal{F}_{in}\sigma_i^+a_n+\text{H.c.}).
\end{equation}
A reasonable choice for the intensity fluctuations is to consider an Ornstein-Uhlenbeck process, and set its mean and variance  to
\begin{equation}
\overline{\Delta\Omega_{\rm d}(t)}=0,\hspace{2ex}\overline{\Delta\Omega_{\rm d}(t)\Delta\Omega_{\rm d}(t)}=\textstyle{\frac{c\tau}{2}}=\zeta_{\rm I}^2\Omega_{\rm d}^2,
\end{equation}
such that $\zeta_{\rm I}$ fixes the relative intensity fluctuations, which we vary in the range  $\zeta_{\rm I}\approx10^{-4}$. In addition, we assume that the correlation time of the intensity fluctuations is longer than the gate time, and set it to $\tau=1\,$ms. 

\begin{figure}
\centering
\includegraphics[width=0.9\columnwidth]{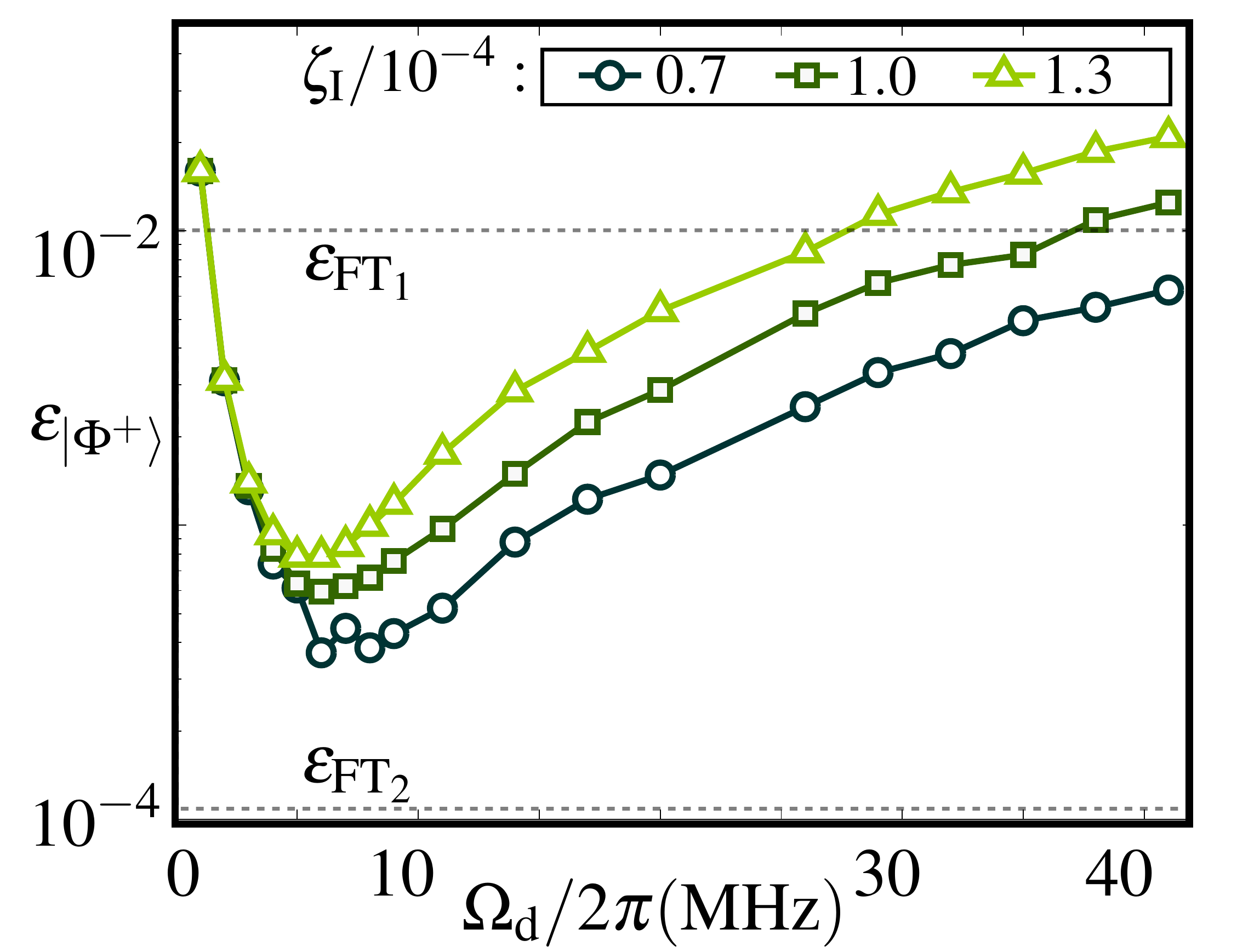}
\caption{ {\bf Intensity noise:} Error in the generation of the Bell state $\ket{\Phi^-}$ from the initial state $\ket{00}\otimes\ket{\bar{n}_1,\bar{n}_2}$ with $\bar{n}_n=0$, for a time evolution including the  noise in the intensity of the microwave driving. The results shown include a statistical average over $N_{\rm n}=10^3$ realizations of the noise process per point, and we set a truncation of the vibrational Hilbert spaces to hold a maximum of $n_{\rm max}=7$ phonons per mode. The error is shown as a function of the applied microwave driving strength, and for different values of the relative noise intensity $\zeta \in \{0.7,1.0,1.3\}10^{-4}$. Gate errors  below the fault-tolerance threshold (FT) can be achieved $\epsilon_{2{\rm q}}<\epsilon_{\rm FT_1}$ for driving strengths $\Omega_{\rm d}/2\pi\approx 7\,$MHz.  }
\label{fig_intensity_noise}
\end{figure}

In Fig.~\ref{fig_intensity_noise}, we integrate numerically the Hamiltonian with a fluctuating microwave intensity~\eqref{dss_intensity_noise} { and a refocusing spin-echo pulse at half the gate time}. As seen in this figure, for intensity stabilization within the $\zeta\approx10^{-4}$ regime, and intermediate drivings $\Omega_{\rm d}/2\pi\in[5,10]$ MHz, the gate error is still well below the higher threshold $\epsilon_{\rm FT_1}$, and approaches the fault-tolerance threshold $\epsilon_{\rm FT_2}$ for optimal drivings and the smallest noise $\zeta_{\rm I}\approx 0.7\cdot 10^{-4}$. In any case, however, the full advantage of the strong-driving limit shown in Figs.~\ref{fig_thermal} and~\ref{fig_dephasing_noise}, where $\epsilon_{2{\rm q}}<\epsilon_{\rm FT_2}$ is attained for sufficiently strong microwave intensities, is lost due to the associated increase of the intensity fluctuations. Hence, it would be desirable to modify the scheme such that it becomes more robust to intensity fluctuations, while preserving its resilience to thermal, dephasing, and phase noise. This is accomplished in the following section. 

 Let us point out that the specific form of the curve in Fig.~\ref{fig_intensity_noise} depends on our  choice of parameters. Still, it also describes qualitatively the behavior of the gate under driving-intensity noise for other parameters. Let us recall that  the microwave driving  suppresses the unwanted contributions of the forces in the $\sigma^y$ and $\sigma^z$ bases, while we 
minimise  the detrimental effect of the spin-dependent force in $\sigma^x$ by closing the phase space trajectories. Thus, we expect the gate error to decrease with increasing driving. Yet, by increasing the driving strength, the noise will eventually become some percentage of the effective qubit-qubit coupling $J_{12}^{\rm dss}$. Since the noise acts in the same basis as the gate, it will eventually start to deteriorate the gate performance outweighing the benefits of the driving. Hence, we expect a minimum error at a certain driving
after which the error increases with stronger drivings. Now, if we make faster gates, the qubit-qubit coupling $J_{12}^{\rm dss}$ is larger, and we expect the minimum  error to appear for a stronger driving. This behavior is confirmed by Fig.~\ref{fig_intensity_noise_fast_gates}, where we carried out a numerical simulation of a faster gate with equal noise parameters, and found that the error minimum is shifted to larger driving strengths. The same considerations can be applied for slower gates, where the minimum would arise at slower drivings.

\subsection{Doubly-driven geometric phase gates}
Inspired by recent results for prolonging  single-qubit coherence times by means of continuous drivings~\cite{concatenated_dd}, we present a modification of the driven single-sideband Hamiltonian~\eqref{dss}, which will allow us to obtain two-qubit gates that are also  robust to intensity fluctuations of the microwave driving. We complement the qubit Hamiltonian~\eqref{qubit} with a secondary microwave driving, such that $ H_{\rm q}\to H_{\rm q,2}$, where
\begin{equation}
\label{intensity_noise_qubit}
 H_{\rm q,2}=\sum_{i=1}^N\half\omega_{0}\sigma_i^z+\half(\Omega_{\rm d}(t)\sigma_i^+\ee^{-\ii\omega_{\rm d}t}+\tilde{\Omega}_{\rm d}(t)\sigma_i^+\ee^{-\ii\tilde{\omega}_{\rm d}t}+\text{H.c.}),
 \end{equation}
and the secondary driving is characterized by a fluctuating Rabi frequency $\tilde{\Omega}_{\rm d}(t)=\tilde{\Omega}_{\rm d}+\Delta\tilde{\Omega}_{\rm d}(t)$,  and a frequency $\tilde{\omega}_{\rm d}$, fulfilling $\tilde{\omega}_{\rm d }\approx \omega_0$, and $\tilde{\Omega}_{\rm d}\ll\omega_0$, where we have assumed again that $\tilde{\Omega}_{\rm d}\in\mathbb{R}$ and $\Delta\tilde{\Omega}_{\rm d}(t)$ is an Ornstein-Uhlenbeck process. This yields the {\it doubly-driven single-sideband Hamiltonian}
\begin{equation}
\label{two_tone}
H_{\rm 2dss}=H_{\rm q,2}+H_{\rm p}+H_{\rm qp},
\end{equation}
where the phonon $H_{\rm p}$ and qubit-phonon $H_{\rm qp}$ terms correspond to Eqs.~\eqref{phonons} and~\eqref{qubit_phonon} described above.

 We show below that this doubly-driven  model 
leads to a geometric phase gate with an additional property: it is robust to intensity fluctuations of the first microwave driving $\Delta\Omega_{\rm d}(t)$, while being sensitive to intensity fluctuations of the secondary driving $\Delta\tilde{\Omega}_{\rm d}(t)$.  This already tells us that the secondary driving should be much weaker  $\tilde{\Omega}_{\rm d}\ll{\Omega}_{\rm d}$, such that the impact of its intensity fluctuations on the gate performance is considerably weaker. However, in order to find a more detailed parameter regime for the high-fidelity gates, we need to explore Eq.~\eqref{two_tone} analytically.

Based on the experience gained  from the analytical study of the single-driving geometric phase gates (see Sec.~\ref{magnus}),
we move to the "dressed-state" interaction picture with respect to the first driving~\eqref{dressed_basis}, which is again assumed to be on resonance with the qubit $\omega_{\rm d}=\omega_0$, and $\Omega_{\rm d}\in\mathbb{R}$. By focusing first on the qubit part of the Hamiltonian, which parallels the discussion in~\cite{concatenated_dd}, we get the following expression
\begin{equation}
\label{two_tone_ip}
\begin{split}
\tilde{H}_{\rm q}\!&=\sum_i\frac{\Delta\Omega_{\rm d}(t)}{2}\sigma_i^x+\\
+&\sum_i \left( \frac{\tilde{\Omega}_{\rm d}(t)}{4}(\!\sigma_i^x+\ii\sigma_i^y\cos(\Omega_{\rm d}t)-\ii\sigma_i^z\sin(\Omega_{\rm d}t)\!) \ee^{-\ii \tilde{\delta}_{\rm d} t} + {\rm H.c.} \right).
\end{split}
\end{equation}
where we have introduced the detuning of the secondary driving $\tilde{\delta}_{\rm d}=\tilde{\omega}_{\rm d}-\omega_0$. If this detuning is set correctly, the secondary driving will decouple the qubit from the intensity fluctuations in the first line of the above equation, just as the dephasing noise is minimized by the first driving (see Eq.~\eqref{eq:par3.2} in Sec.~\ref{dephasing_section}). In particular, this is achieved for $\tilde{\delta_{\rm d}}=\Omega_{\rm d}$, such that a rotating-wave approximation for $\tilde{\Omega}_{\rm d}\ll 4 \Omega_{\rm d}$ leads to
\begin{equation}
\label{z_driving}
\tilde{H}_{\rm q}=\half\sum_i\Delta\Omega_{\rm d}(t)\sigma_i^x-\fourth\sum_i (\tilde{\Omega}_{\rm d}+\Delta\tilde{\Omega}_{\rm d}(t))\sigma_i^z.
\end{equation}
If we now move to the dressed-state interaction picture of the secondary driving, which we shall call the double dressed-state interaction picture
\begin{equation}
\label{double_dressed_basis}
\tilde{\tilde{U}}_{\rm d}(t)=\ee^{-\ii t\sum_i\fourth\tilde{\Omega}_{\rm d}\sigma_i^z}\ee^{\ii t\sum_i\half\Omega_{\rm d}\sigma_i^x}\ee^{\ii t \sum_i\half\omega_0\sigma_i^z}\ee^{\ii t\sum_n\omega_na_n^{\dagger}a_n},
\end{equation} 
the qubit Hamiltonian can be rewritten as
\begin{equation}
\begin{split}
\tilde{\tilde{H}}_{\rm q}=&\frac{1}{2}\sum_i\Delta\Omega_{\rm d}(t)\big(\cos(\half\tilde{\Omega}_{\rm d}t)\sigma_i^x+\sin(\half\tilde{\Omega}_{\rm d}t)\sigma_i^y\big)-\\
-&\frac{1}{4}\sum_i \Delta\tilde{\Omega}_{\rm d}(t)\sigma_i^z.
\end{split}
\end{equation}
In this picture, it becomes clear that the noisy terms due to the first driving become rapidly rotating even for a weaker second driving provided that $\zeta_{\rm I}\Omega_{\rm d}\ll\tilde{\Omega}_{\rm d}\ll 4\Omega_{\rm d}$. Hence, we can neglect them in a rotating-wave approximation, such that the qubit is only sensitive to the fluctuations of the second driving $\tilde{\tilde{H}}_{\rm q}\approx
-\frac{1}{4}\sum_i \Delta\tilde{\Omega}_{\rm d}(t)\sigma_i^z$, which have a weaker effect.

So far, we have only treated the qubit part of the doubly-driven single-sideband Hamiltonian~\eqref{two_tone}. The crucial point to address now is whether the two strong drivings can be combined with the qubit-phonon couplings responsible for the entangling gate. This question is by no means trivial, since the secondary driving acting in the $\sigma^z$-basis~\eqref{z_driving} will make the $\sigma^x$ state-dependent force~\eqref{x_force} rotate very fast, and can thus inhibit the required qubit-phonon couplings. To find the correct parameter regime, let us express the  qubit-phonon Hamiltonian in the double dressed-state interaction picture
\begin{equation}
\label{single_sideband_dip}
\begin{split}
\tilde{\tilde{H}}_{\rm qp}\!=\!\sum \limits_{i,n}\! \frac{\mathcal{F}_{in}}{2} \big(\!\sigma_i^xf_x(t)+\sigma_i^yf_y(t)-\ii\sigma_i^zf_z(t)\!\big) a_n \ee^{-\ii \delta_n t} + {\rm H.c.},
\end{split}
\end{equation}
where we have introduced the following time-dependences
\begin{equation}
\label{time_dependences}
\begin{split}
f_x(t)&=\cos(\half\tilde{\Omega}_{\rm d}t)-\ii\sin(\half\tilde{\Omega}_{\rm d}t)\cos({\Omega}_{\rm d}t),\\
f_y(t)&=\sin(\half\tilde{\Omega}_{\rm d}t)+\ii\cos(\half\tilde{\Omega}_{\rm d}t)\cos({\Omega}_{\rm d}t),\\
f_z(t)&=\sin({\Omega}_{\rm d}t).\\
\end{split}
\end{equation}
We now study all the possibilities. {\it (i)} If the laser-induced detunings fulfil $\delta_n\ll\tilde{\Omega}_{\rm d}\ll\Omega_{\rm d}$, then  all the terms in~\eqref{single_sideband_dip} become rapidly rotating, and the qubit-phonon coupling is inhibited by the double driving. {\it (ii)} If $\delta_n\approx\tilde{\Omega}_{\rm d}\ll\Omega_{\rm d}$, such that $\half|\mathcal{F}_{in}|\ll\tilde{\Omega}_{\rm d}\ll\Omega_{\rm d}$, then only the $\sigma^z$ force can be neglected in a rotating-wave approximation. Unfortunately, the $\sigma^x$ and $\sigma^y$ forces then contribute equally, and we would lose the geometric character of the entangling gate (i.e. it will be very sensitive to thermal motion of the ions). {\it (iii)} Finally, if $\tilde{\Omega}_{\rm d}\ll\delta_n\approx\Omega_{\rm d}$, such that $\half|\mathcal{F}_{in}|\ll\tilde{\Omega}_{\rm d},\:\Omega_{\rm d}$, then both $\sigma^{x/y}$ forces become rapidly rotating and can be neglected, but the $\sigma^z$ force preserves its nearly-resonant character. To achieve this condition, instead of setting the laser frequency close to the bare sideband $\omega_{\rm L}\approx \omega_0 - \omega_n$ (see Sec.~\ref{hardware}), we need to adjust it close to the dressed sideband resonance
\begin{equation}
\label{double_detunings}
\omega_{\rm L}\approx\omega_0-(\omega_n-\Omega_{\rm d}), \hspace{2ex}\tilde{\delta}_n=\omega_{\rm L}-(\omega_0+\Omega_{\rm d}-\omega_n).
\end{equation} 
In this regime, we obtain a single state-dependent force 
\begin{equation}
\label{single_sideband_z}
\begin{split}
\tilde{\tilde{H}}_{\rm qp}\!\approx-\!\sum \limits_{i,n}\! \frac{\mathcal{F}_{in}}{4} \sigma_i^z a_n \ee^{-\ii \tilde{\delta}_n t} + {\rm H.c.},
\end{split}
\end{equation}
 which will lead to a geometric phase gate in a different basis. We also note that the condition for the laser Rabi frequency is modified in this case to $|\Omega_{\rm L}|\ll 8 (\omega_n-|\tilde{\Omega}_{\rm d}/2|)$.
 
 %%%%%%%%%%%%%%%%%%%%%%%%%%
\begin{figure}
\centering
\includegraphics[width=0.75\columnwidth]{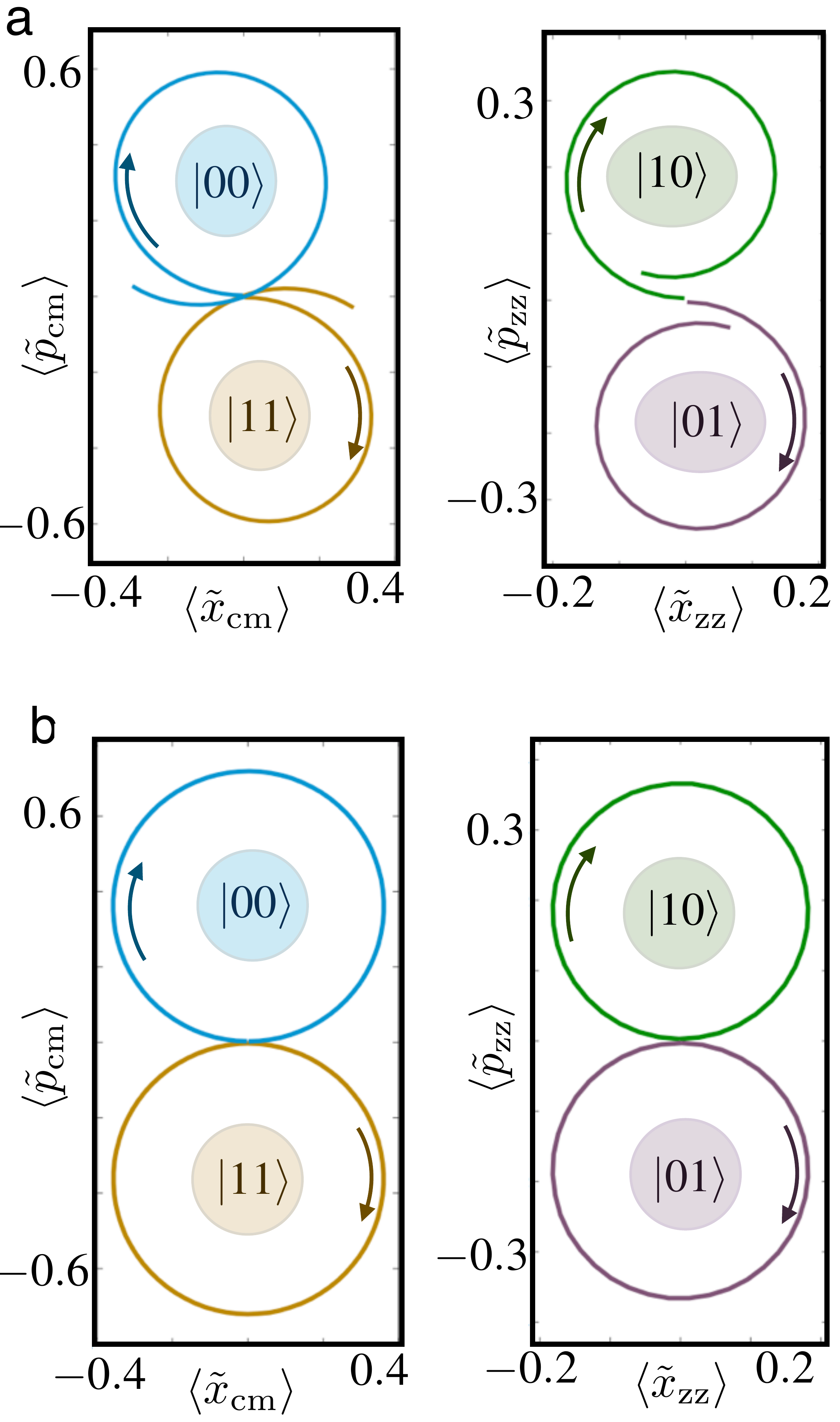}
\caption{{\bf Doubly-driven phase-space trajectories:} Numerical analysis of the phase-space trajectories  under the doubly-driven  single-sideband Hamiltonian~\eqref{single_sideband_dip}. {\bf (a)} Trajectories of the forced normal modes for a weak secondary driving  $\tilde{\Omega}_{\rm d}/2\pi=10\,$kHz, and a strong primary driving $\Omega_{\rm d}/2\pi=10\,$ MHz. Both of the phonon modes are in the vacuum state. In the left panel, the initial spin state is $\ket{11}$ or $\ket{00}$, and according to the forces~\eqref{forces}, only the center-of-mass mode is driven in phase space $\langle\tilde{x}_{\rm cm}\rangle=1/\sqrt{2}\langle a_1+a_1^{\dagger}\rangle,\langle\tilde{p}_{\rm cm}\rangle=\ii/\sqrt{2}\langle a_1^{\dagger}- a_1\rangle$. In the right panel, the initial spin state is $\ket{10}$ or $\ket{01}$, such that only the zigzag mode develops a driven trajectory in this case $\langle\tilde{x}_{\rm zz}\rangle=1/\sqrt{2}\langle a_2+a_2^{\dagger}\rangle,\langle\tilde{p}_{\rm zz}\rangle=\ii/\sqrt{2}\langle a_2^{\dagger}- a_2\rangle$. In both cases, the trajectories are not closed. {\bf (b)} Same as {\bf (a)} but in the regime where the secondary driving is also strong $\tilde{\Omega}_{\rm d}/2\pi=6\,$MHz. As an effect of the strong driving, the trajectories are closed, and the gate inherits a geometric character. In the numerics we have set a truncation of $n_{\rm max} = 5$ to the vibrational Hilbert spaces.}
\label{trajectories_z}
\end{figure}
%%%%%%%%%%%%%%%%%%%%%%%%%%

Let us now support the above discussion by integrating numerically the dynamics induced by the doubly-driven single-sideband Hamiltonian in Eqs.~\eqref{single_sideband_dip}-\eqref{time_dependences}, such that the laser is now tuned to the regime~\eqref{double_detunings}. From the above discussion, we know that by applying a sufficiently-strong secondary driving $\tilde{\Omega}_{\rm d}\gg|\mathcal{F}_{in}|$,  we are left with a single state-dependent force in the $\sigma^z$-basis with a halved strength $\half\mathcal{F}_{in}\to \fourth \mathcal{F}_{in}$ with respect to the previous $\sigma^x$-force~\eqref{x_force}. In this regime, we should recover circular phase-space trajectories when the initial state of the qubits is an eigenstate of the $\sigma_1^z\sigma_2^z$ operator. In Fig.~\ref{trajectories_z},  we analyze the strong-driving dynamics numerically. For weak secondary drivings [Fig.~\ref{trajectories_z}{\bf (a)}], we observe how the different phase-space trajectories are not closed. Conversely, for strong secondary drivings [Fig.~\ref{trajectories_z}{\bf (b)}], we recover circular orbits that close exactly at the expected gate time. These numerical results support the validity of our arguments leading to Eq.~\eqref{single_sideband_z}.

%%%%%%%%%%%%%%%%%%%%%%%%%%
\begin{figure}
\centering
\includegraphics[width=0.75\columnwidth]{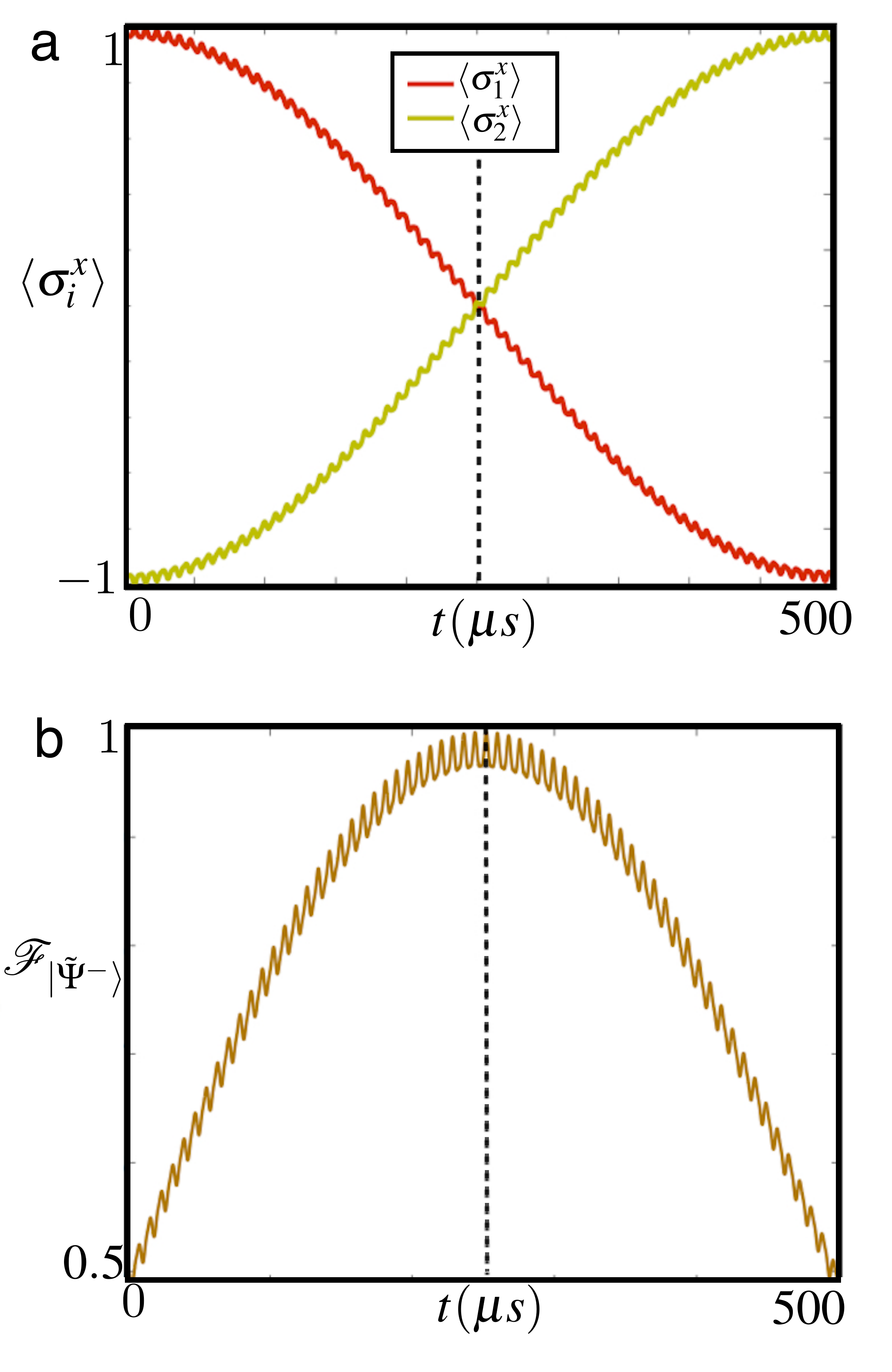}
\caption{{\bf Entangling doubly-driven geometric phase gate:} We display the dynamics by  numerical integration of the unitary evolution induced by~\eqref{single_sideband_dip}. We consider an initial state $\ket{\psi_0}=\ket{+-}\otimes\ket{\bar{n}_1,\bar{n}_2}$, where $\bar{n}_1=\bar{n}_2= 0$ are the initial mean number of phonons for the cm and zz modes, and $\ket{\pm}=(\ket{0}\pm\ket{1})/\sqrt{2}$. We have to set a truncation $n_{\rm max}=5$ to the vibrational Hilbert spaces.   In {\bf (a)}, we represent the dynamics of local qubit operators $\langle \sigma_i^x\rangle$, which show the qubit flip $\ket{+-}\to\ket{-+}$ after $t\approx 500\mu$s.  In {\bf (b)}, we display the fidelity between the time-evolved state  and the Bell state $\ket{\tilde{\Psi}^-}=(\ket{+-}-\ii\ket{-+})/\sqrt{2}$. At $t\approx 250\mu$s, this fidelity approaches unity, which supports the validity of the doubly-driven single-sideband two-qubit gates.}
\label{z_gate}
\end{figure}
%%%%%%%%%%%%%%%%%%%%%%%%%%

Starting from the state-dependent force in the $\sigma^z$-basis~\eqref{single_sideband_z}, we can apply once more the Magnus expansion to get the following approximation to the time-evolution operator
\begin{equation}
\label{eff_zz}
U_{\rm z}(t_{\rm g})\approx\tilde{\tilde{U}}^{\dagger}_{\rm d}(t)\ee^{-\ii H_{\rm ddss}t_{\rm g}},\hspace{2ex}
 H_{\rm ddss}=\sum_{i,j}J^{\rm{ddss}}_{ij}\sigma_i^z\sigma_j^z,
\end{equation}
where the coupling strengths now are given by
\begin{equation}
J_{ij}^{\rm ddss}=-\sum_n \frac{1}{16\tilde{\delta_n}}\mathcal{F}_{in} \mathcal{F}_{jn}^{*}.
\end{equation}
Note that , in order to obtain this time evolution, it is necessary to neglect all other contributions coming from the Magnus expansion. This is possible by considering the strong-driving limit $\half|\mathcal{F}_{in}|\ll\tilde{\Omega}_{\rm d}\ll 4\Omega_{\rm d}$, where additional conditions similar to the single-driving case~\eqref{constraints} are also fulfilled
\begin{equation}
\label{constraints_double_driving}
t_{\rm g}= r \frac{2 \pi}{\tilde{\delta}_1}, \hspace{1ex} \tilde{\delta}_2=k \tilde{\delta}_1, \hspace{1ex}\Omega_{\rm d}=p \tilde{\delta}_1,\hspace{1ex}\tilde{\Omega}_{\rm d}=q \tilde{\delta}_1, \hspace{2ex}\: r,k,p,q\in \mathbb{Z}.
\end{equation}
These conditions allow us to close exactly the phase-space trajectories [Fig.~\ref{trajectories_z}{\bf (b)}] or, equivalently, to minimize the residual qubit-phonon couplings that would make the gate sensitive to the thermal motion of the ions. We have found that these conditions can be met by imposing
\begin{equation}
|\Omega_{\rm L}|=\frac{2|\tilde{\delta}_1|}{\eta_1 \sqrt{\frac{r}{2}\left(1-\frac{1}{2 \xi} \right)}},
\label{dd_laser_intensity}
\end{equation}
in analogy to our older choice~\eqref{eq:par2.6} for the single-driving gates. We use the same values for the detunings as in Table~\ref {tab_1} with $\delta_{1/2} \to \tilde{\delta}_{1/2}$, but now set the parameter $r=32$ to optimize the gate fidelities, at the price of diminishing the gate speed.  Note that the unitary evolution~\eqref{eff_zz} leads to the following table for $t_{\rm g}=\pi/(8J_{12}^{\rm ddss})$
\begin{equation}
\label{table_bis2}
\begin{split}
\textstyle{\ket{++}\to\ket{\tilde{\Phi}^-}=\frac{1}{\sqrt{2}}(\ket{++}-\ii\sgn(J^{\rm{ddss}}_{12})\ket{--}),}\\
\textstyle{\ket{+-}\to\ket{\tilde{\Psi}^-}=\frac{1}{\sqrt{2}}(\ket{+-}-\ii\sgn(J^{\rm{ddss}}_{12})\ket{-+}),}\\
\textstyle{\ket{-+}\to\ket{\tilde{\Psi}^+}=\frac{1}{\sqrt{2}}(\ket{+-}+\ii\sgn(J^{\rm{ddss}}_{12})\ket{-+}),}\\
\textstyle{\ket{--}\to\ket{\tilde{\Phi}^+}=\frac{1}{\sqrt{2}}(\ket{++}+\ii\sgn(J^{\rm{ddss}}_{12})\ket{--}).}\\
\end{split}
\end{equation}
In Fig.~\ref{z_gate}{\bf (b)}, we represent the fidelity for the unitary generation of the entangled state $\ket{\tilde{\Psi}^-}$ from the initially unentangled state $\ket{+-}$. It becomes clear from this figure, that the fidelities that can be achieved are again close to 100$\%$. 

However, the fast and small-amplitude oscillations in the fidelity set an upper limit to the achieved fidelities, which is below the desired fault-tolerance thresholds. In order to improve the gate fidelities, we introduce a simple spin-echo refocusing pulse, such that the complete time-evolution operator is 
$U_{\rm full} (t_{\rm g},0)= U_{\rm z,noise}(t_{\rm g},\frac{t_{\rm g}}{2})(\sigma^y_1 \sigma_2^y) U_{\rm z,noise}(\frac{t_{\rm g}}{2},0)$. In this expression, we have also included the intensity fluctuations of the first driving, such that  $U_{\rm z,noise}(t_2,t_1)$ is the time-evolution operator induced by the noisy Hamiltonian
\begin{equation}
\begin{split}
H_{\rm z,noise}=&\frac{1}{2}\sum_i\Delta\Omega_{\rm d}(t)\big(\cos(\half\tilde{\Omega}_{\rm d}t)\sigma_i^x+\sin(\half\tilde{\Omega}_{\rm d}t)\sigma_i^y\big)\\
+&\sum \limits_{i,n}\! \frac{\mathcal{F}_{in}}{2} \big(\!\sigma_i^xf_x(t)+\sigma_i^yf_y(t)-\ii\sigma_i^zf_z(t)\!\big) a_n \ee^{-\ii \delta_n t} + {\rm H.c.},
\end{split}
\end{equation}
where all the parameters have already been introduced above. Note that, due to the refocusing pulse, the target entangled state obtained from $\ket{+-}$ is no longer $\ket{\tilde{\Psi}^-}$, but rather $\ket{\tilde{\Psi}^+}$.

\begin{figure}
\centering
\includegraphics[width=0.9\columnwidth]{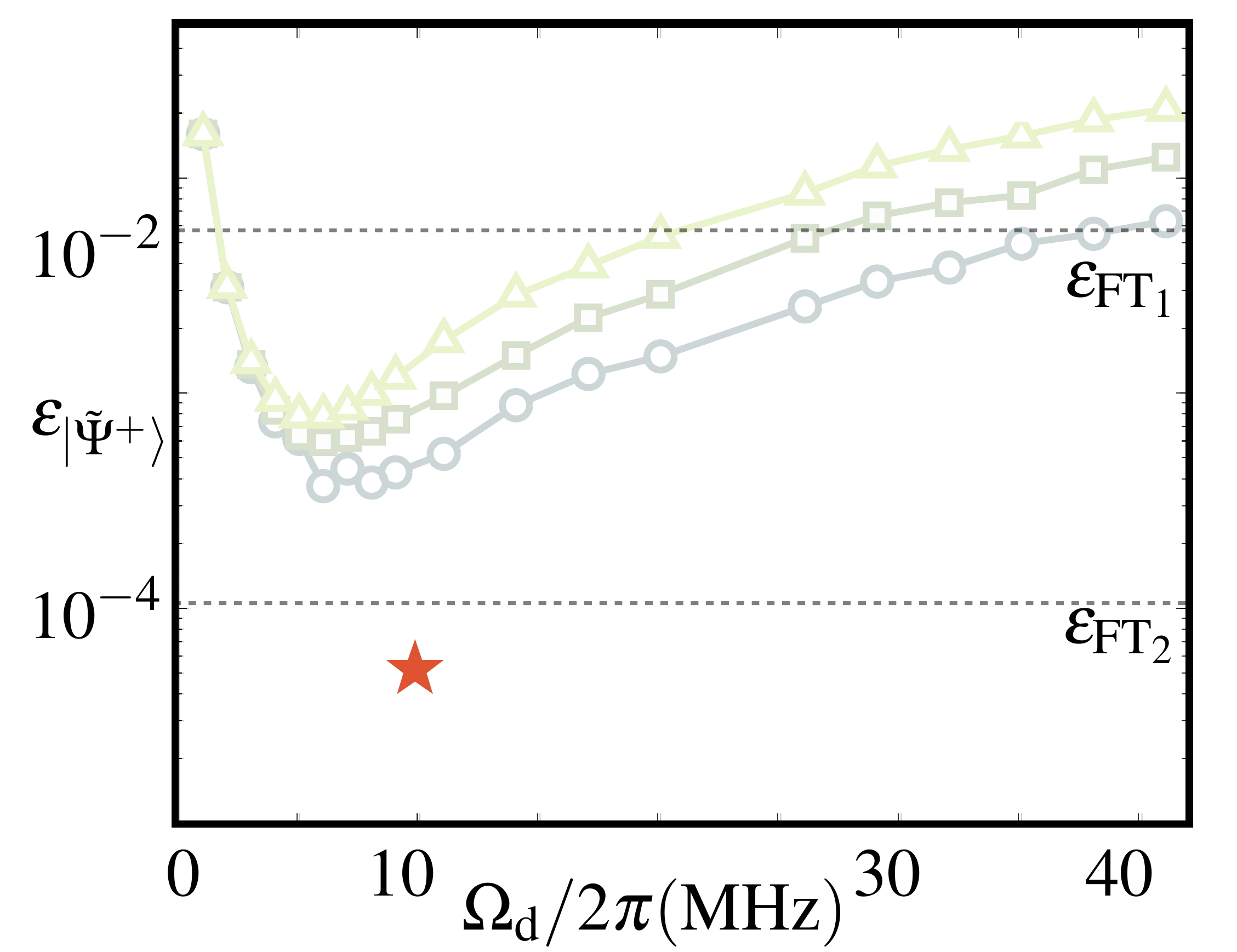}
\caption{ {\bf Intensity noise in the doubly-driven geometric phase gate:} Error in the generation of the Bell state $\ket{\tilde{\Psi}^+}$ from the initial state $\ket{++}\otimes\ket{\bar{n}_1,\bar{n}_2}$ with $\bar{n}_n = 0$, for a time evolution including the  noise in the intensity of the first microwave driving of strength $\Omega_{\rm d}/2\pi=10\,$MHz, and relative noise of $\zeta_{\rm I}=10^{-4}$. We consider a secondary driving with $\tilde{\Omega}_{\rm d}/2\pi=6\,$MHz and no intensity fluctuations. The results shown include a statistical average over $N_{\rm n}=10^3$ realizations of the noise process, and set a truncation of the vibrational Hilbert spaces to hold a maximum of $n_{\rm max}=5$ phonons per mode. The obtained error for the doubly-driven gate is represented as a red star, which is well below the second threshold $\epsilon_{2{\rm q}}<\epsilon_{\rm FT_2}$. The older errors for the single microwave driving [Fig.~\ref{fig_intensity_noise}] are also included in the background for comparison.}
\label{fig_intensity_noise_dd}
\end{figure}

In Fig.~\ref{fig_intensity_noise_dd}, we represent the achieved error of this doubly-driven geometric phase gate for the generation of the entangled state $\ket{\tilde{\Psi}^+}$ in the presence of noise on the first driving. We note that with
our choice of parameters for the simulation displayed in Fig.~\ref{fig_intensity_noise_dd} we do not fulfil $|\tilde{\Omega}_{\rm d}|\ll|\Omega_{\rm d}|$ but the analytic calculation showed that it is enough to fulfil 
$|\tilde{\Omega}_{\rm d}|\ll4|\Omega_{\rm d}|$ to obtain the doubly-driven gate in the $\sigma^z$ basis. The second, more relaxed constraint is met here.

\begin{figure}[hbtp]
\centering
\includegraphics[width=0.9\columnwidth]{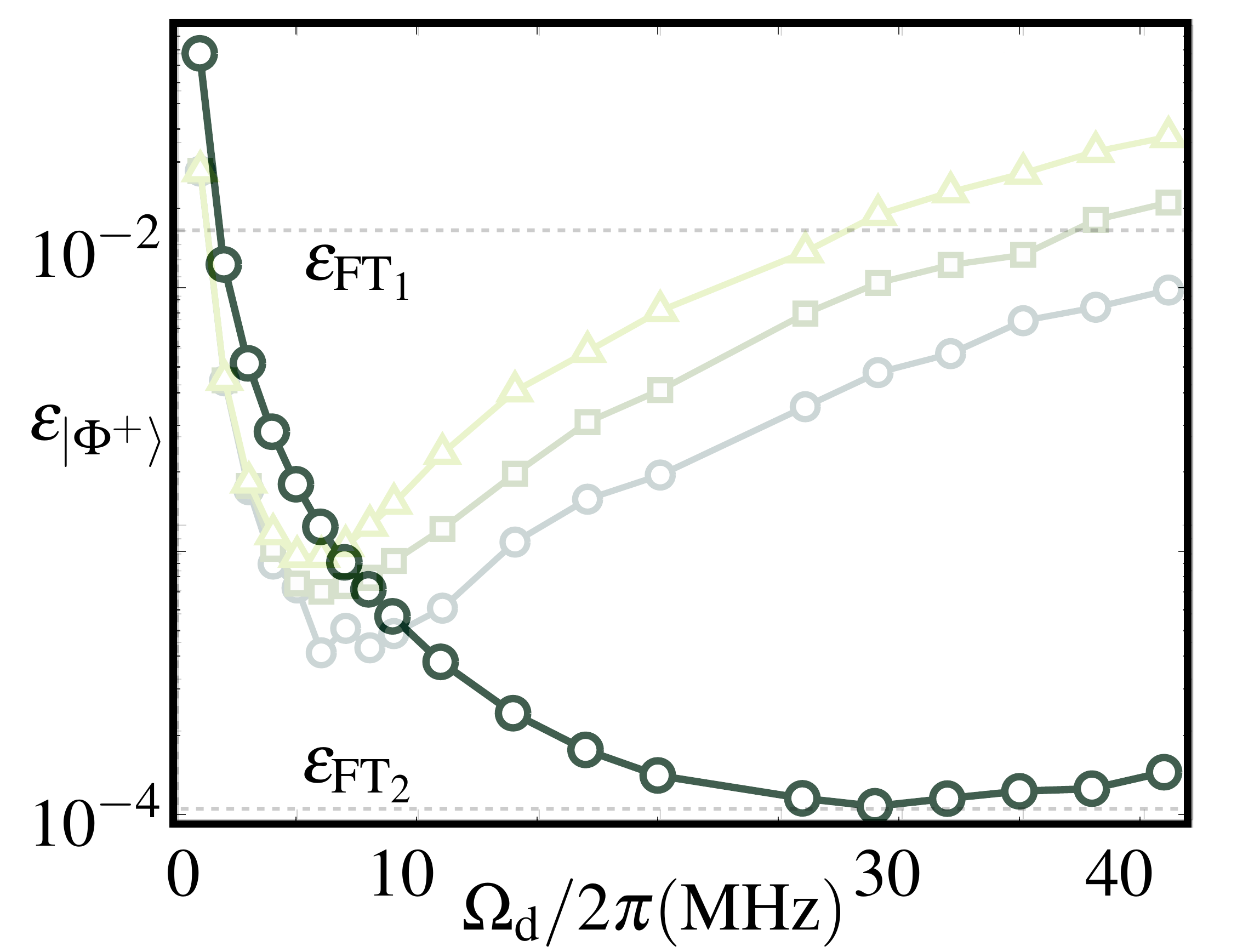}
\caption{ {\bf Intensity noise for a faster gate}: Error in the generation of the Bell state $\ket{\Phi^-}$ from the initial state $\ket{00}\otimes\ket{\bar{n}_1,\bar{n}_2}$ with $\bar{n}_n = 0$, for a time evolution including the  noise in the intensity of the microwave driving with a relative noise of $\zeta_{\rm I}=0.7 \cdot 10^{-4}$. We set the parameter $r$ from eq.~\eqref{constraints} to $r=2$ which corresponds to a gate time $t_{\rm g}\approx 16\,\mu$s. The results shown include a statistical average over $N_{\rm n}=10^3$ realizations of the noise process, and we truncated the vibrational Hilbert spaces at a maximum phonon number $n_{\rm max}=7$ phonons per mode. The gate error is displayed as a function of the applied microwave driving strength. Gate errors of $\epsilon_{\rm 2q} \approx 1\cdot 10^{-4}$ can be achieved.}
\label{fig_intensity_noise_fast_gates}
\end{figure}

This numerical result displays the superior performance of this new gate, and its resilience to the intensity fluctuations of the first microwave driving allowing for $\epsilon_{2{\rm q}}<\epsilon_{{\rm FT}_2}$. 
{ At this point, it might be worth emphasizing the benefit of the secondary driving. Noting that the scheme also works for clock states, one might ask if the impact of the intensity fluctuations of the microwave driving
could be reduced by making faster gates. In fact, assuming a clock state, we can increase the gate speed by decreasing the parameter $r$ from Eq.~\eqref{constraints} without compromising the error due to the dephasing noise. In Fig.~\ref{fig_intensity_noise_fast_gates}, the error for a gate with $r=2$ in the presence of intensity noise $\zeta_I=0.7\cdot 10^{-4}$  on the microwave driving is shown. As can be seen in the figure, the error reaches $\epsilon_{2q} \approx 1\cdot 10^{-4}$,  while the error achieved with the secondary driving is  lower  $\epsilon_{2q}\approx 5\cdot 10^{-5}$.}

 We remark that intensity fluctuations of the secondary driving have not been considered. Such fluctuations will eventually be the limiting factor for the gate fidelity. However, since the secondary driving is weaker than the first one, their effect will be smaller. Moreover, due to the excellent fidelity obtained, it may also be possible to reduce the second driving further, making the gate more insensitive to its fluctuations. Another possibility would be to introduce a tertiary microwave driving, which would be now detuned by the Rabi frequency of the secondary one $\tilde{\tilde{\omega}}_{\rm d}-\omega_0=\tilde{\Omega}_{\rm d}$. This driving would act as a decoupling mechanism from intensity fluctuations of the second driving. We believe it should be possible to modify the sideband resonance~\eqref{double_detunings}, such that we obtain a triply-driven geometric phase gate in a different basis (possibly the $\sigma^y$ basis) that enjoys a further robustness against all these sources of noise. Such  concatenated schemes can be followed until the desired fidelities are achieved.

\section{Conclusions and Outlook}
\label{conc}
In this manuscript, we have demonstrated theoretically a two-qubit entangling gate for trapped ions, which relies only on a single red-sideband excitation and a strong microwave driving tuned to the carrier transition.
By properly choosing the laser and microwave frequencies, we have shown analytically that  the controlled dynamics corresponds to a geometric phase gate in the $\sigma_x$ basis, which we have called the {\it driven geometric phase gate}. We have shown numerically that the gate is able to overcome the imperfections associated to  thermal, dephasing, 
and phase noise, while achieving gate speeds comparable to state-of-the-art implementations. In particular, we have shown    that such a driven geometric phase gate can attain errors well below the fault-tolerance threshold $\epsilon_{{\rm FT}_2}\sim10^{-4}$ for sufficiently-strong microwave drivings. 

We have also analyzed numerically how this gate is still sensitive to intensity fluctuations of the microwave driving,
which might be the currently limiting factor to its  accuracy. To overcome this drawback, we have devised a new scheme with a weaker secondary microwave driving, which makes the gate robust against fluctuations in the
first driving intensity. We have showed both analytically and numerically that, by setting the laser and microwave parameters in a certain regime,  the dynamics correspond to  a geometric phase gate in the $\sigma_z$ basis. In this case, the limiting factor for the gate fidelity will be the  fluctuations in the secondary microwave driving.

{\it Acknowledgements.--}  We acknowledge very useful correspondence with D. Leibfried, which has helped us in clarifying the interrelation between the original proposal~\cite{ss_gate}, the experimental realization~\cite{nist_ss_gate}, and the results presented in this manuscript. This work was supported by  PICC and  by the Alexander von Humboldt Foundation. A.B. thanks FIS2009-10061, and QUITEMAD.

\appendix

\section{Magnus expansion for the driven single-sideband Hamiltonian}\label{app1}

The Magnus expansion (ME) allows us to write the time evolution operator  of a system with a time-dependent Hamiltonian $H(t)$ as $U(t,t_0)=\ee^{\Omega(t,t_0)}$, where $\Omega(t,t_0)$ is an anti-hermitian operator that can be written in a perturbative series $\Omega(t,t_0)=\sum_{k=1}^{\infty} \Omega_k(t,t_0)$~\cite{me_ref}. 
In this work, we are interested in the ME to second order, where the first- and second-order terms are given by
\begin{align}
\Omega_1(t,t_0) &= -\ii\int_{t_0}^{t}\dd t_1 H(t_1) 
\label{eq:app2} \\
\Omega_2(t,t_0) & = -\frac{1}{2}\int_{t_0}^{t} \dd t_1 \int_{t_0}^{t_1} \dd t_2 [H(t_1),H(t_2)].  
\label{eq:app3}
\end{align}
In Sec.~\ref{magnus}, we used the results for the second-order  ME for the qubit-phonon Hamiltonian~\eqref{single_sideband_ip2}, namely
\begin{equation}
\tilde{H}_{\rm qp}\!=\!\sum \limits_{j,n}\! \frac{\mathcal{F}_{jn}}{2} (\!\sigma_j^x+\ii\sigma_j^y\cos(\Omega_{\rm d}t)-\ii\sigma_j^z\sin(\Omega_{\rm d}t)\!) a_n \ee^{-\ii \delta_n t} + {\rm H.c.},
\end{equation}
which is expressed in an interaction picture with respect to the microwave driving. In this Appendix, we present a detailed derivation of the ME and discuss its more relevant terms.

Let us start by considering the first-order contribution, which is calculated directly from integrals of the qubit-phonon Hamiltonian, and  leads to the following result
\begin{equation}
\begin{split}
\Omega_1 & (t,0) = \sum \limits_{j,n} \frac{\mathcal{F}_{jn}}{2} \left[ \left( \ee^{-\ii \delta_n t} -1 \right) \frac{1}{\delta_n} \sigma_j^x a_n + \right.  \\
  &+ \left( \ee^{\ii (\Omega_{\rm d} - \delta_n) t} - 1 \right) \frac{1}{2(\Omega_{\rm d} -\delta_n)} (-\ii \sigma_j^y a_n + \sigma_j^z a_n)  \\
	& + \left. \left(\ee^{-\ii (\Omega_{\rm d} + \delta_n) t} - 1 \right) \frac{1}{2(\Omega_{\rm d} +\delta_n)} (\ii \sigma_j^y a_n + \sigma_j^z a_n)  \right] -{\rm H.c.},
\end{split}
\label{eq:app4a}
\end{equation}
which coincides with Eq.~\eqref{Omega1} in the main text, and contains the different state-dependent forces.
The second-order term  can be split into three different parts, namely
\begin{equation}
\Omega_2(t,0) = \Omega_2^a(t,0) + \Omega_2^b(t,0) + \Omega_2^c(t,0).  
\label{eq:app4b}
\end{equation}
The first part $\Omega_2^a(t,0)$ contains the terms that are linear in time, and can be interpreted as a spin-spin Hamiltonian generated by the three non-commuting state-dependent forces 
\begin{equation}
\begin{split}
\Omega_2^a (t,0) = -\ii t \sum \limits_{j,k}&\left(  \fourth{J^{\rm{eff}}_{jk}} \sigma_j^x \sigma_k^x +\half\sum_n\Delta \Omega_{jn} \sigma_j^x \delta_{jk}+   \right.  \\
	 & \left. +\textstyle{\frac{1}{8}} M^{\rm{eff}}_{jk} \sigma_j^y \sigma_k^z+ \frac{1}{16} K^{\rm{eff}}_{jk} (\sigma_j^y \sigma_k^y + \sigma_j^z \sigma_k^z) \right),
\end{split}
\label{eq:app4c}
\end{equation}
where $\delta_{jk}$ is the Kronecker delta and the coupling constants $J^{\rm{eff}}_{jk}$, $M^{\rm{eff}}_{jk}$, $K^{\rm{eff}}_{jk}$ and the driving corrections $\Delta \Omega_{jn}$ are given by
\begin{equation}
\begin{split}
	J^{\rm{eff}}_{jk} & = -\sum \limits_{n}\frac{1}{\delta_n} {\rm Re}\{\mathcal{F}_{jn} \mathcal{F}_{kn}^{*}\}, \\
	K^{\rm{eff}}_{jk} & =  \sum \limits_{n}\left( \frac{1}{\Omega_{\rm d} - \delta_n} - \frac{1}{\Omega_{\rm d} + \delta_n} \right){\rm Re}\{ \mathcal{F}_{jn} \mathcal{F}_{kn}^{*}\} , \\
	M^{\rm{eff}}_{jk} & =  \sum \limits_{n}\left( \frac{1}{\Omega_{\rm d} - \delta_n} + \frac{1}{\Omega_{\rm d} + \delta_n} \right) {\rm Im}\{\mathcal{F}_{jn} \mathcal{F}_{kn}^{*}\} ,\\
	\Delta \Omega_{jn} & = -  \frac{1}{4} \left( \frac{1}{\Omega_{\rm d} - \delta_n} + \frac{1}{\Omega_{\rm d} + \delta_n} \right)  |\mathcal{F}_{jn}|^2. 
\end{split}
\label{eq:app4d}
\end{equation}
At this point, let us note that our specific configuration considers a propagation direction of the laser beams fulfilling ${\bf k}_{\rm L}\cdot{\bf r}_j^0 = 0$, such that the forces $\mathcal{F}_{jn}$   are purely imaginary (see Eq.~\eqref{forces}), and the couplings $M_{jk}^{\rm eff}$  vanish exactly.
Moreover,  in the strong-driving limit $\delta_n \ll \Omega_{\rm d} $, we find that $K^{\rm{eff}}_{jk} \propto \mathcal{F}_{jn} \mathcal{F}_{kn}^{*} /\Omega_{\rm d}^2$, which can be neglected to order $\mathcal{O}(\xi)$, with $\xi=(\Omega_{\rm L}\eta_{n})^2/\Omega_{\rm d}^2\ll1$, as was done in Eq.~\eqref{Omega2} of the main text. Altogether, we conclude that the most important part  is
\begin{equation}
\label{a_leading}
\Omega_2^a (t) \approx -\ii t \sum \limits_{j,k}\big(\fourth{J^{\rm{eff}}_{jk}} \sigma_j^x \sigma_k^x +\half \sum_n \Delta \Omega_{jn} \sigma_j^x \delta_{jk}  \big). 
\end{equation}

Let us now move to the second part $\Omega_2^b(t,0)$, which contains oscillatory couplings between the internal qubit states
\begin{equation}
\begin{split}
\Omega_2^b & (t,0) = \ii \sum \limits_{j} \frac{J^{\rm{eff}}_{jj}}{4 \Omega_{\rm d}} \left(  \sin(\Omega_{\rm d} t) \sigma_j^z + \left(1 - \cos(\Omega_{\rm d} t) \right) \sigma_j^y \right) -\\
&	- \sum \limits_{j,k,n} \left[ \frac{\mathcal{F}_{jn}\mathcal{F}_{kn}^{*}}{8} \left( \hat{a}^1_{jkn}  \big( \ee^{-\ii \delta_n t} -1 \big) 
+  \right. \right. \\
& \left. \left. +\hat{a}^2_{jkn}\big(  \ee^{\ii (\Omega_d - \delta_n) t} -1 \big) + \hat{a}^3_{jkn} \big( \ee^{-\ii (\Omega_d + \delta_n) t} -1 \big)   \right) -{\rm H.c.} \right],
\end{split}
\label{eq:app4e}
\end{equation}
where we have introduced the two-body spin operators
\begin{equation}
\begin{split}
\hat{a}^1_{jkn}  &=  -\frac{1}{\delta_n^2} \sigma_j^x \sigma_k^x  + \frac{1}{2 \delta_n (\Omega_{\rm d} + \delta_n)} (+\ii \sigma_j^x \sigma_k^y - \sigma_j^x \sigma_k^z)  \\
 & \hspace{12ex}+ \frac{1}{2 \delta_n (\Omega_{\rm d} - \delta_n)} (-\ii \sigma_j^x \sigma_k^y - \sigma_j^x \sigma_k^z), \\
\hat{a}^2_{jkn} &=  \frac{1}{2 \delta_n (\Omega_{\rm d} - \delta_n)} (+\ii \sigma_j^y \sigma_k^x - \sigma_j^z \sigma_k^x), \\
\hat{a}^3_{jkn} &=  \frac{1}{2 \delta_n (\Omega_{\rm d} + \delta_n)} (- \ii \sigma_j^y \sigma_k^x + \sigma_j^z \sigma_k^x). 
\end{split}
\label{eq:app4f}
\end{equation}
Let us now discuss which are the leading terms in the above expression~\eqref{eq:app4e}. The first line of Eq.~\eqref{eq:app4e} only contributes perturbatively since $J_{jk}^{\rm eff}\ll8\Omega_{\rm d}$. From the remaining contributions to~\eqref{eq:app4e}, note that most of the terms in~\eqref{eq:app4f} scale with $\mathcal{F}_{jn} \mathcal{F}_{kn}^{*} /\Omega_{\rm d}\delta_n$, and can be thus neglected to order $\mathcal{O}(\chi)$, where $\chi=(\Omega_{\rm L}\eta_{n})^2/\Omega_{\rm d}\delta_n\ll1$, as was done in Eq.~\eqref{Omega2} of the main text. Finally, considering again that the forces are purely real in our setup, we get the following leading contribution
\begin{equation}
\label{b_leading}
\Omega_2^b (t,0) \approx -\ii  \sum \limits_{j,k}\frac{\mathcal{F}_{jn}\mathcal{F}_{kn}^*\sin(\delta_nt)}{4\delta_n^2} \sigma_j^x \sigma_k^x. 
\end{equation}

The third and final  part $\Omega_2^c (t)$ contains contains the residual spin-phonon couplings, and reads as follows
\begin{equation}
\begin{split}
  \Omega_2^c (t)& =  \ii t \sum \limits_{j,n}\Delta\Omega_{jn} \sigma_j^x a_n^{\dagger} a_n +  \sum \limits_{j,m \neq n}  (\hat{f}_{1j}^{nm}(t)a_m^{\dagger} a_n-\text{H.c.})\\
 & +  \sum \limits_{j,m,n} \left( \hat{f}_{2j}^{nm} (t) a_m a_n - \hat{f}_{3j}^{nm} (t) a^{\dagger}_m a_n -{\rm H.c.}\right) ,\\
\end{split}
\label{eq:app4g}
\end{equation}
where $\Delta\Omega_{jn}$ was introduced in~\eqref{eq:app4d}, and we have introduced the following time-dependent single-qubit operators
\begin{equation}
\label{f1}
\hat{f}_{1j}^{nm}(t)=\frac{\mathcal{F}_{jn}\mathcal{F}_{jm}^*}{8 (\delta_n - \delta_m)}\!\! \left( \frac{1}{\Omega_{\rm d} - \delta_m} + \frac{1}{\Omega_{\rm d} + \delta_m}\right)\!\!\big(\ee^{-\ii(\delta_n-\delta_m)t} - 1 \big)\sigma_j^x,
\end{equation}
together with
\begin{equation}
\begin{split}
 \hat{f}_{2j}^{nm} (t)& = b_{jmn}^1 \big(\ee^{-\ii \delta_n t} -1 \big) + b_{jmn}^2 \big(\ee^{-\ii ( \delta_n + \delta_m) t} -1 \big) \\
	& + b_{jmn}^3 \big(\ee^{\ii (\Omega_{\rm d} - \delta_n) t} - 1 \big) + b_{jmn}^4 \big(\ee^{-\ii (\Omega_{\rm d} + \delta_n) t} - 1 \big) \\
	& + b_{jmn}^5 \big(\ee^{\ii (\Omega_{\rm d} - \delta_n - \delta_m) t} - 1 \big) + b_{jmn}^6 \big(\ee^{-\ii (\Omega_{\rm d} + \delta_n + \delta_m) t} - 1 \big), 
\label{eq:app4h}
\end{split}
\end{equation}
and, finally,
\begin{equation}
\begin{split}
 \hat{f}_{3j}^{nm} (t) & = c_{jmn}^1 \big(\ee^{-\ii \delta_n t} -1 \big) \\
   &+c_{jmn}^2 \big(\ee^{\ii (\Omega_d - \delta_n) t} - 1 \big) + c_{jmn}^3 \big(\ee^{-\ii (\Omega_d + \delta_n) t} - 1 \big)  \\
	&+ c_{jmn}^4 \big(\ee^{\ii (\Omega_d + \delta_m - \delta_n) t} - 1 \big) + c_{jmn}^5 \big(\ee^{-\ii (\Omega_d + \delta_n - \delta_m) t} - 1 \big). 
\label{eq:app4j}
\end{split}
\end{equation}
In these last two expressions, for notational convenience, we have introduced the following 
list of single-qubit operators 
\begin{equation}
\begin{split}
b_{jmn}^1 &= \frac{\mathcal{F}_{jn} \mathcal{F}_{jm}^{*}}{8\delta_n (\Omega_{\rm d} - \delta_m)} (\ii \sigma_j^y-\sigma_j^z)  + \frac{\mathcal{F}_{jn} \mathcal{F}_{jm}^{*}}{8\delta_n (\Omega_{\rm d} + \delta_m)} (\sigma_j^z +\ii \sigma_j^y), \\
b_{jmn}^2 &= -\frac{\mathcal{F}_{jn} \mathcal{F}_{jm}^{*}}{8(\delta_n + \delta_m) } \left( \frac{1}{\Omega_{\rm d} - \delta_m} + \frac{1}{\Omega_d + \delta_m} \right) \sigma_j^x, \\
b_{jmn}^3 &= \frac{\mathcal{F}_{jn} \mathcal{F}_{jm}^{*}}{8\delta_m (\Omega_{\rm d} - \delta_n)} (\sigma_j^z-\ii \sigma_j^y  ),  \\
b_{jmn}^4 &= \frac{\mathcal{F}_{jn} \mathcal{F}_{jm}^{*}}{8\delta_m (\Omega_{\rm d} + \delta_n)} (-\sigma_j^z-\ii \sigma_j^y ), \\
b_{jmn}^5 &=\frac{\mathcal{F}_{jn} \mathcal{F}_{jm}^{*}}{8\delta_m (\Omega_{\rm d} - \delta_n - \delta_m)} (-\sigma_j^z+\ii \sigma_j^y ), \\
b_{jmn}^6 &= \frac{\mathcal{F}_{jn} \mathcal{F}_{jm}^{*}}{8\delta_m (\Omega_{\rm d} + \delta_n + \delta_m)} ( \sigma_j^z+\ii \sigma_j^y ), 
\end{split}
\label{eq:app4i}
\end{equation}
and also
\begin{equation}
\begin{split}
c_{jmn}^1 &= \frac{\mathcal{F}_{jn} \mathcal{F}_{jm}^{*}}{8\delta_n (\Omega_{\rm d} + \delta_m)} (\ii \sigma_j^y-\sigma_j^z )  + \frac{\mathcal{F}_{jn} \mathcal{F}_{jm}^{*}}{8\delta_n (\Omega_{\rm d} - \delta_m)} ( \ii \sigma_j^y+\sigma_j^z), \\
c_{jmn}^2 &= \frac{\mathcal{F}_{jn} \mathcal{F}_{jm}^{*}}{8\delta_m (\Omega_{\rm d} + \delta_n)} (-\sigma_j^z+\ii \sigma_j^y ),  \\
c_{jmn}^3 &= \frac{\mathcal{F}_{jn} \mathcal{F}_{jm}^{*}}{8\delta_m (\Omega_{\rm d} - \delta_n)} ( \sigma_j^z+\ii \sigma_j^y ), \\
c_{jmn}^4 &= \frac{\mathcal{F}_{jn} \mathcal{F}_{jm}^{*}}{8\delta_m (\Omega_{\rm d} + \delta_m - \delta_n)} ( -\sigma_j^z+\ii \sigma_j^y ), \\
c_{jmn}^5 &= \frac{\mathcal{F}_{jn} \mathcal{F}_{jm}^{*}}{8\delta_m (\Omega_{\rm d} + \delta_n - \delta_m)} (\sigma_j^z+\ii \sigma_j^y ). 
\end{split}
\label{eq:app4j}
\end{equation}
Once all the expressions of the Magnus expansion have been described, let us analyze the leading-order contribution to the dynamics. First of all, note that in the strong-driving limit $\Omega_{\rm d}\gg\delta_n$, all the expressions in Eqs.~\eqref{eq:app4i} and~\eqref{eq:app4j} scale as $\mathcal{F}_{jn} \mathcal{F}_{jm}^{*} /\Omega_{\rm d}\delta_n$, which is again on the order of $\mathcal{O}(\chi)$, where $\chi=(\Omega_{\rm L}\eta_{n})^2/\Omega_{\rm d}\delta_n\ll1$, and can be directly neglected. Hence, the leading contribution to this part is 
\begin{equation}
\label{c_leading}
\Omega_2^c (t)\approx  \ii t \sum \limits_{j,n}\Delta\Omega_{jn} \sigma_j^x a_n^{\dagger} a_n +  \sum \limits_{j,m \neq n}  (\hat{f}_{1j}^{nm}(t)a_m^{\dagger} a_n-\text{H.c.}).
\end{equation}
It is now easy to convince oneself that all the leading contributions to the second-order~\eqref{a_leading},~\eqref{b_leading} and~\eqref{c_leading}, form the expression~\eqref{Omega2} used in the main text.

\section{Stochastic processes for the noise sources}
\label{app2}

In this part of the appendix, we present a more detailed discussion of the origin of the dephasing, phase, and intensity noise in ion trap setups. As described in the main text, all these noise sources can be modeled by a particular stochastic Hamiltonian. We use this appendix to introduce the particular stochastic models used in this work, and their main properties.

We use two continuous memoryless stochastic processes, also known as Markov processes, namely the Ornstein-Uhlenbeck and the Wiener process. In the following, we introduce briefly both  processes, and present some of their properties that will be important for our numerical simulations. Subsequently, we will describe how they are used to model the fluctuations of a specific
quantity in the trapped-ion setup.

{\it a) Ornstein-Uhlenbeck process.--}
The Ornstein-Uhlenbeck (O-U) process $O(t)$ is determined by a diffusion constant $c$ and a correlation time $\tau$, and evolves according to the following stochastic differential equation
\begin{equation}
\frac{\text{d} O (t)}{\text{d} t} = -\frac{1}{\tau} O (t) + \sqrt{c} \Gamma(t),
\label{eq:app5b}
\end{equation}
where $\Gamma (t)$ is a Gaussian white noise. Remarkably, equation~\eqref{eq:app5b} is exactly solvable~\cite{ouproc}, which allows to show that the O-U process is a normal process with mean and variance
\begin{equation}
\overline{O (t)} = O (0) \ee^{-t/\tau}, \hspace{3ex}
\text{Var}\{O  (t)\} = \frac{c \tau}{2} \left( 1- \ee^{-2t/\tau} \right).
\label{eq:app5c}
\end{equation}
As one can see in \eqref{eq:app5c}, the correlation time $\tau$ sets the time scale on which the asymptotic values of the process are reached. Assuming a zero mean, 
the auto-covariance of the O-U process is given by
\begin{equation}
\overline{O (t) O (0)} = \frac{c \tau}{2} \ee^{-t/\tau} (1-\ee^{-2t/\tau}),
\label{eq:app5d}
\end{equation}
where one can see that $\tau$ also governs the time-scale over which the noise is correlated. Finally, the essential ingredient for the numerical simulations is the
exact update formula 
\begin{equation}
O (t + \Delta t) = O (t) \ee^{-\Delta t/\tau} + \left[ \frac{c \tau}{2} (1-\ee^{- 2 \Delta t/\tau} )\right]^{1/2} n,
\label{eq:app5e}
\end{equation}
where $n$ is a unit Gaussian random variable. We note that this formula is valid for any finite time step $\Delta t$, which will turn out to be very useful for the numerical simulations.

{\it b) Wiener process.--}
In the limit of very large correlation times $\tau \to \infty$, an O-U process with diffusion constant $c$ becomes another type of Markov process, which is known in the literature as a driftless Wiener process $W(t)$~\cite{ouproc}. 
The standard form of the Langevin equation for the Wiener process $W(t)$ reads as follows
\begin{equation}
W(t+\Delta t) = W(t) + (c\Delta t)^{1/2}n,
\label{w1}
\end{equation}
where $n$ is again a unit Gaussian random variable. Note that this serves again as an update formula that will be very useful for our numerical simulations. Introducing once more the Gaussian white noise function $\Gamma (t)$, equation~ \eqref{w1} can be recast as a stochastic differential equation 
\begin{equation}
\frac{{\rm d} W (t)}{\rm{d}t} = \sqrt{c} \Gamma(t),
\label{w2}
\end{equation}
which corresponds formally to Eq. \eqref{eq:app5b} in the limit $\tau \to \infty$.
The Wiener process is also a Gaussian process and is characterized by a mean and variance
\begin{equation}
\label{w3}
\overline{W (t)} = W (0),\hspace{3ex}\text{Var}\{W (t) \} = c t.
\end{equation}

In the following paragraphs, we will discuss the concrete application of the above  stochastic processes for the simulations of {\it dephasing}, {\it phase}, and {\it intensity} noise. For the numerical simulations, the values of $O(t),W(t)$ are sampled according to the update formulas in  Eqs. \eqref{eq:app5e} and~\eqref{w1}. Note that for all simulations we set $O(0) = W(0)= 0$ for the stochastic processes.

\vspace{1ex}
{\it i) Dephasing noise.--} The internal structure of the atomic ions can be perturbed  by uncontrolled electric or magnetic fields. Magnetic-field fluctuations are typically leading, and are expected to be the major source of dephasing \cite{wineland_review}, which ultimately leads to the loss of
coherence. To understand the effect of  a fluctuating magnetic field, note that the Zeeman shift  changes the  qubit resonance frequency by an amount $\Delta \omega_0 $ given by

\begin{equation}
\Delta \omega_0 = \partial_{B} \omega_0 |_{B_0} (B - B_0) + \frac{1}{2} \partial^2_{B} \omega_0 |_{B_0} (B - B_0)^2,
\label{eq:dpn0}
\end{equation}
where the field $B$ fluctuates around the average value $B_0$. Many hyperfine transitions are first-order field insensitive at $B=B_0= 0$, but the need to resolve the hyperfine
structure for the definition of our qubit (see Fig.~\ref{fig_1}) requires a finite external magnetic field. Interestingly, there is a family of states/transitions, often referred to as clock states~\cite{clock_states}, where the linear Zeeman shift vanishes at a certain value $B_0$, yielding thus a higher resilience to magnetic-field noise. 

Although our scheme applies equally well to such clock states, we would like to consider magnetic-field sensitive states, and show that the strong driving also protects the qubits from magnetic-field noise. Therefore, we will assume that $\Delta\omega_0(t)=-g\mu_{\rm B}B(t)$, where $g$ is the hyperfine g-factor, $\mu_{\rm B}$ the Bohr magneton, and $B(t)$ a randomly fluctuating magnetic field. Thus, $\Delta\omega_0(t)$ also fluctuates randomly, and the noise is described by a stochastic Hamiltonian term
\begin{equation}
H_{\rm fluc}=\half\sum_i\Delta\omega_0(t)\sigma_i^z,
\end{equation}
as introduced in Eq.~\eqref{eq:par3.1} of the main text. For clock states, $\Delta\omega_0(t)$ will contain the weaker quadratic Zeeman shift, and the ac-Stark shifts due to fluctuating laser intensities.

%%%%%%%%%%%%%%
\begin{figure}
\centering
\includegraphics[width=0.8\columnwidth]{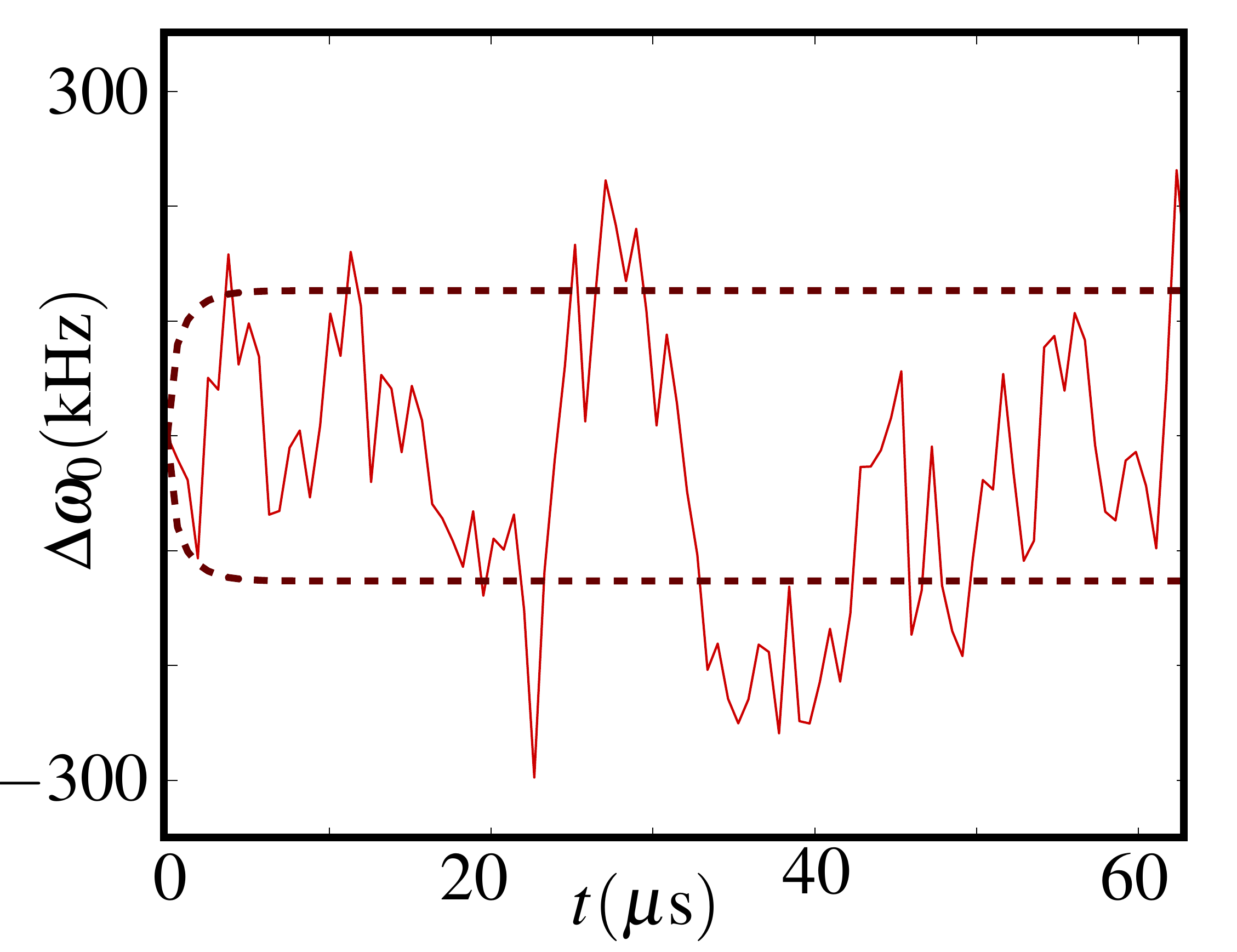}
\caption{ {\bf Dephasing noise process:} Fluctuations of the qubit resonance frequency $\Delta\omega_0(t)$ during a single realization of the driven geometric phase gate $t_{\rm g}\approx 63\,\mu$s. The dashed lines represent the standard deviation that follows from~\eqref{eq:dpn1} of the O-U process that models the dephasing, while the  solid line represents the stochastic process itself.  We set  the decoherence time to $T_2=25\,\mu$s, which fixes the correlation time  $\tau=2.5\,\mu$s, and the diffusion constant $c=2/(T_2\tau^2)$ of  the O-U process.}
\label{dephasing_noise_process}
\end{figure}
%%%%%%%%%%%%%%%%%

In order to reproduce the exponential decay of the coherences typically observed in experiments, we modeled the fluctuations of the resonance frequency $\Delta \omega_0 (t)$ 
as an Ornstein-Uhlenbeck (O-U) process~\eqref{eq:app5b}. According to Eqs. \eqref{eq:app5c} the fluctuations in the resonance frequency are characterized by
\begin{equation}
\overline{\Delta \omega_0 (t)} = 0 , \hspace{3ex}
\text{Var}\{\Delta \omega_0 (t)\} = \overline{\Delta \omega_0^2 (t)} = \frac{c \tau}{2} \left( 1- \ee^{-2t/\tau} \right),
\label{eq:dpn1}
\end{equation}
(recall that we set $\Delta \omega (0)=0$) and following \eqref{eq:app5e} the update formula reads
\begin{equation}
\Delta \omega_0 (t + \Delta t) = \Delta \omega_0 (t) \ee^{-\Delta t/\tau} + \left[ \frac{c \tau}{2} (1-\ee^{- 2 \Delta t/\tau}) \right]^{1/2} n.
\label{eq:dpn2}
\end{equation}

In our noise model, the temporal decay of the qubit coherences is given by 
\begin{equation}
\langle \sigma_i^x (t_{\rm f})\rangle  = \langle \sigma_i^x (0) \rangle \ee^{- \half \langle Y^2 (t_{\rm f}) \rangle},
\label{eq:dpn3}
\end{equation}
where $Y(t_{\rm f})$ is the integral of the O-U process~\cite{ouproc}.
The exact expression for $\langle Y^2 (t_{\rm f})\rangle$ is given by 
\begin{equation}
\label{eq:app6a}
\begin{split}
\langle Y^2 (t_{\rm f}) \rangle &= \int \limits_{0}^{t_{\rm f}} \rm{d}t' \int \limits_{0}^{t_{\rm f}} \rm{d}t'' \langle \Delta\omega_0(t') \Delta\omega_0(t'')\rangle \\
&=  c\tau^2 \left( t_{\rm f}- 2 \tau (1-\ee^{-t_{\rm f}/\tau}) +\frac{\tau}{2} (1-\ee^{-2t_{\rm f}/\tau}) \right).
\end{split}
\end{equation}
In the limit $t_{\rm f} \gg \tau$, and using Eq.~\eqref{eq:app6a}, we get $\langle \sigma_i^x(t_{\rm f})  \rangle = \langle \sigma_i^x(0)  \rangle \ee^{- t_{\rm f}/T_2}$,  where we have introduced the dephasing time 
\begin{equation}
T_2=\frac{2}{c \tau^2}.
\label{eq:app6b}
\end{equation}
For all numerical simulations, we set $\tau=0.1 T_2$. Thus, after choosing $T_2$ the positive constant $c$ is determined by Eq. \eqref{eq:app6b}. As we shall show in the following, it is required that 
the time step $\Delta t$ used in the numerical simulations be much smaller than $\tau$.

%%%%%%%%%%%%%%%%%%%%%%%%%%%%%%%
\begin{figure}
\centering
\includegraphics[width=0.8\columnwidth]{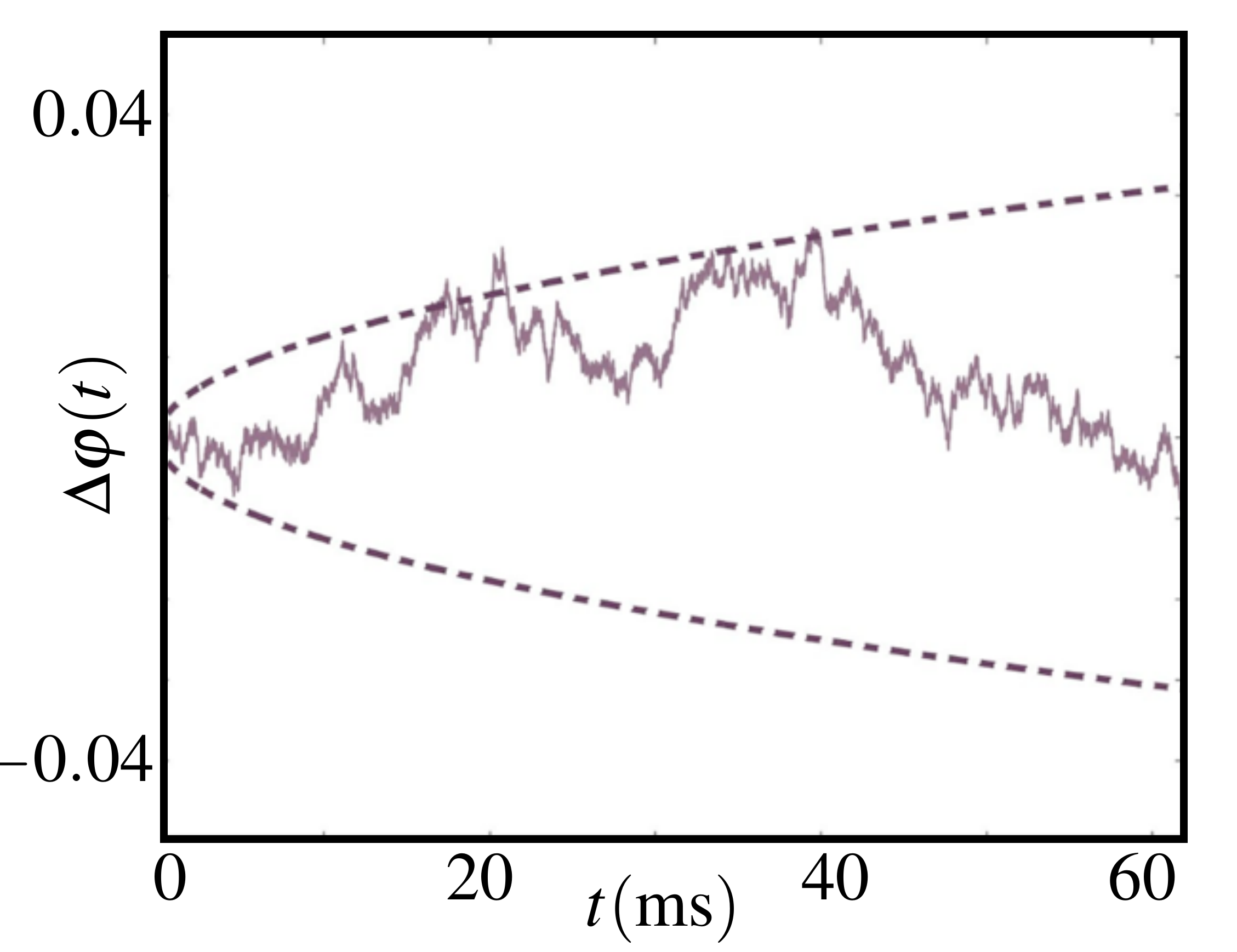}
\caption{ {\bf Phase noise process:}  Fluctuations of the laser phases $\Delta\varphi(t)$ leading to a slow drift during subsequent repetitions of the driven geometric phase gate, $t_{\rm f}=10^3t_{\rm g}\approx 63\,$ms. The dashed lines represent the standard deviation that follows from~\eqref{phase_noise_w} of the Wiener process that models the phase drifts, while the  solid line represents the stochastic process itself.  We set  the relative phase fluctuations to be $\zeta_{\rm p}=10^{-2}$, which fixes the diffusion constant of the process according to Eq.~\eqref {eq:pn1}. } 
\label{phase_noise_process}
\end{figure}
%%%%%%%%%%%%%%%%%%%%%%%%%%%%%%%%

Since  time is discretized in the numerical simulations, the integral is approximated by a sum
\begin{equation}
\label{eq:app7}
\begin{split}
\langle Y^2 (t_{\rm f}) \rangle  \approx \: & \frac{c \tau}{2} \sum \limits_{n=1}^{N} \Delta t \ee^{\frac{-n \Delta t}{\tau}} \sum \limits_{n'=1}^{n} \Delta t \ee^{\frac{n' \Delta t}{\tau}} \\
& + \frac{c \tau}{2} \sum \limits_{n=1}^{N} \Delta t \ee^{\frac{n \Delta t}{\tau}} \sum \limits_{n'=n+1}^{N} \Delta t \ee^{\frac{-n' \Delta t}{\tau}},
\end{split}
\end{equation}
where $N \Delta t= t_{\rm f}$. After some  algebra, equation~\eqref{eq:app7} gives
\begin{equation}
\label{eq:app8}
\begin{split}
\langle Y^2 (t_{\rm f}) \rangle \approx &\frac{c \tau}{2} \Delta t^2  \left(\frac{N \ee^{-\Delta t/\tau}}{1-\ee^{-\Delta t/\tau}} - \frac{N\ee^{\Delta t/\tau}}{1-\ee^{ \Delta t/\tau}} \right)\\
 +&\frac{c \tau}{2} \Delta t^2 \bigg(\frac{1-\ee^{-N \Delta t/\tau}}{1-\cosh(\Delta t/\tau)}\bigg).
\end{split}
\end{equation}
In the limits $\Delta t \ll \tau$, and $N \Delta t = t_{\rm f} \gg \tau$, we can approximate Eq.~\eqref{eq:app8} by 
\begin{equation}
\label{eq:app9}
\langle Y^2 (t_{\rm f}) \rangle \approx c \tau^2 \left(N \Delta t - \tau \right) \approx c \tau^2 t_{\rm f},
\end{equation}
giving a good approximation to the analytical result in Eq.~\eqref{eq:app6b}. A similar result is obtained for $\Delta t \approx \tau$. On the other hand, for the case $\Delta t \gg \tau$, we get
\begin{equation}
\label{eq:app9}
\langle Y^2 (t_{\rm f}) \rangle \approx \frac{c \tau}{2} \Delta t t_{\rm f}
% c \tau^2 t_{\rm f}\bigg(\frac{\Delta t}{2\tau}\ee^{\Delta t/\tau}\bigg),
\end{equation}
such that the exponential decay of the coherences will be much faster than expected. Consequently, we chose the time step $\Delta t$ in the numerical simulations to be $\Delta t \ll \tau$.

In Fig.~\ref{dephasing_noise_process}, we represent the fluctuations of the qubit resonance frequency given by an O-U process fulfilling the above requirements. The time-scale  in this figure corresponds to a single realization of the driven geometric phase gate, and we typically average over $N_{\rm n}=10^3$ gate realizations.

\begin{figure}
\centering
\includegraphics[width=0.8\columnwidth]{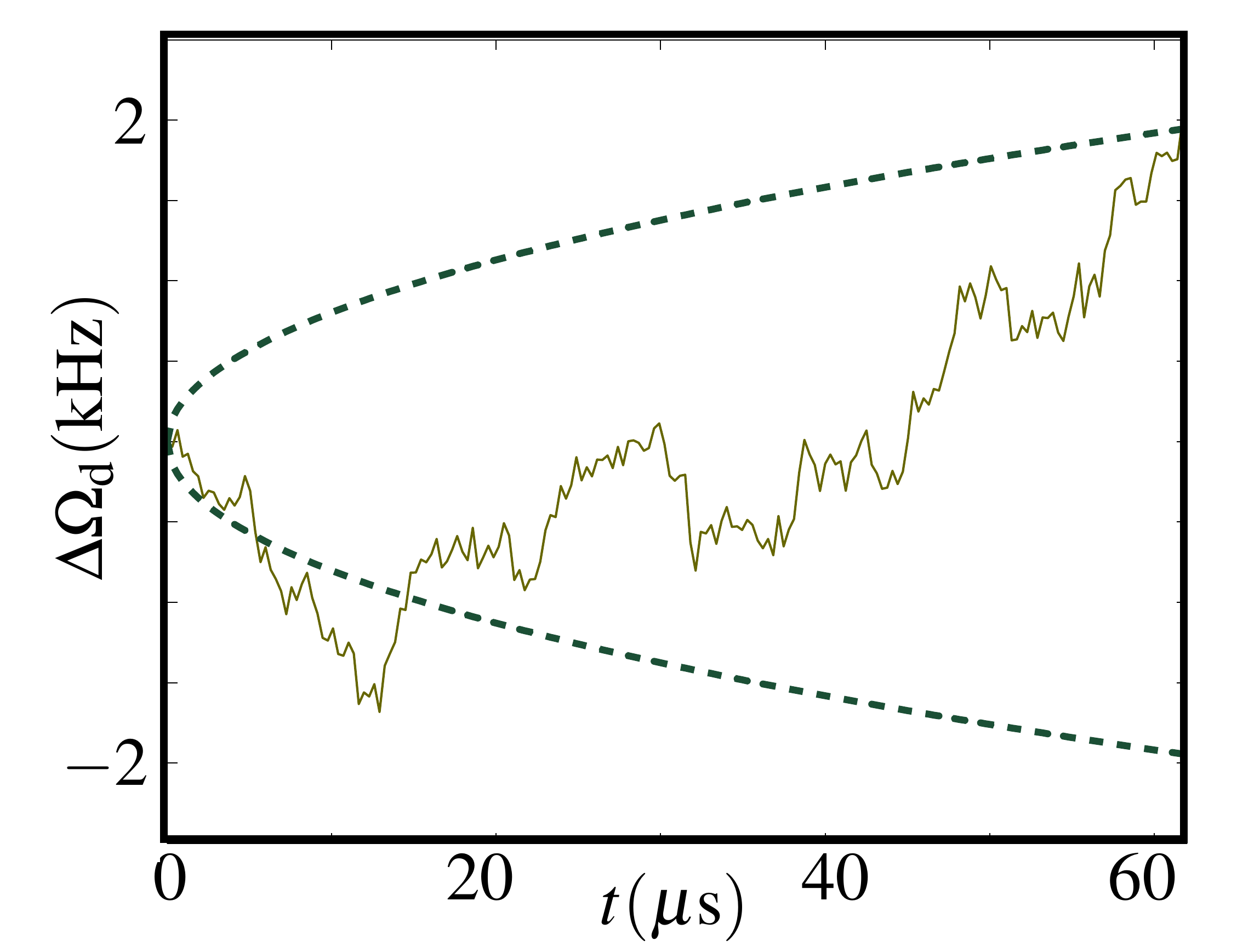}
\caption{ {\bf Intensity noise process:} Fluctuations of the microwave driving intensity $\Delta\Omega_{\rm d}(t)$ during a single realization of the driven geometric phase gate $t_{\rm g}\approx 63\,\mu$s. The dashed lines represent the standard deviation that follows from~\eqref{eq:app5c} of the O-U process that models the intensity fluctuations, while the  solid line represents the stochastic process itself.  We set  the relative intensity fluctuations to be $\zeta_{\rm I}=1.3\cdot10^{-4}$ for a driving of $\Omega_{\rm d} / 2\pi\approx 7\,$MHz, the correlation time was fixed to $\tau=1\,$ms, which fixes the diffusion constant of the process according to Eq.~\eqref{intensity_c}. }
\label{intensity_noise_process}
\end{figure}

\vspace{1ex}
{\it ii) Phase noise.--} Another source of noise in trapped-ion experiments  is due to the fact that the phases of the lasers at the positions of the ions are not constant, but rather  subjected to slow drifts~\cite{phase_noise_haljan}. To study the effects of phase fluctuations on the driven geometric phase gate, we make the substitution $\mathcal{F}_{in}\to \mathcal{F}_{in} \ee^{\ii \Delta\varphi(t)}$ for the laser-induced sideband couplings such that the qubit-phonon couplings~\eqref{raman} become
\begin{equation}
H_{\rm qp}=\sum_{i,n}(\mathcal{F}_{in}\ee^{\ii\Delta\varphi(t)}\sigma_i^+a_n\ee^{-\ii\omega_{\rm L}t}+\text{H.c.}),
\end{equation}
as introduced in Eq.~\eqref{dss_phase_noise} of the main text. As reported in experiments~\cite{exp_phase_noise_schmidt_kaler}, a phase shift of $2 \pi$ takes place on time-scales on the order of $\sim 10\,$s. Since these drifts occur on a very long time-scale, it appears convenient to model these fluctuations as a Wiener process. Therefore, the phase fluctuations evolve according to Eq.~\eqref{w1} as
\begin{equation}
\Delta\varphi(t+\Delta t) = \Delta\varphi(t) + (c\Delta t)^{1/2}n.
\label{phase_noise_update}
\end{equation}
The mean and the variance of the process are given by
\begin{equation}
\label{phase_noise_w}
\overline{\Delta\varphi (t)} = \Delta\varphi(0) = 0,\hspace{3ex} \text{Var} \{\Delta\varphi(t)\} =\overline{\Delta\varphi^2 (t)} = c t,
\end{equation}

In the numerical simulations, we did not rely on the assumption that the phase of the lasers is constant during one gate realization (see the qualitative discussion in the main text). Since the fluctuations during one gate realization are very small, the
relevant  question is if the phase fluctuations affect the gate performance for a large number of consecutive gates. Therefore, we simulated $N_{\rm n}=10^3$ consecutive gate realizations with 
fluctuating sideband couplings $\mathcal{F}_{jn} \ee^{\ii \Delta\varphi(t)}$.

The stochastic process $\Delta\varphi (t)$ describing the phase fluctuations is then fully characterized by its diffusion constant $c$ and its initial value $\Delta \varphi (0)=0$. 
With the results of \cite{exp_phase_noise_schmidt_kaler}, and an expected gate time $t_{\rm g}\approx 63\,\mu$s, we can estimate $c$ following Eqs. \eqref{w3} as
\begin{equation}
c=\frac{(\zeta_{\rm p} \pi)^2}{10^3 t_{\rm g}},\hspace{2ex}\zeta_{\rm p}\approx0.01.
\label{eq:pn1}
\end{equation}
The noise process $\Delta\varphi(t)$ can  then be generated using the update formula~\eqref{phase_noise_update}, which demands the time step $\Delta t$ for the simulation to be sufficiently small. This
requirement should be fulfilled for our particular choice of $N=200$ steps per gate. In Fig.~\ref{phase_noise_process}, we represent the phase fluctuations modeled by a Wiener process fulfilling the above requirements.

\vspace{1ex}
{\it iii) Intensity fluctuations.--} The  scheme for driven geometric phase gates introduced in this paper achieves very high fidelities for entangled two-qubit states in the regime of very high powers of the microwave 
driving. Unfortunately, in this  regime, the output amplitude of microwave sources cannot be held perfectly constant, and  we should  consider fluctuations in the microwave intensity. This leads to a noisy qubit Hamiltonian
 \begin{equation}
 H_{\rm q}=\sum_{i=1}^N\half\omega_{0}\sigma_i^z+\half\big((\Omega_{\rm d}+\Delta\Omega_{\rm d}(t))\sigma_i^+\ee^{-\ii\omega_{\rm d}t}+\text{H.c.}\big),
 \end{equation}
where $\Delta\Omega_{\rm d}(t)$ represents the fluctuations of the microwave Rabi frequencies introduced in Eq.~\eqref{intensity_noise_qubit} of the main text. We model these fluctuations by an O-U process~\eqref{eq:app5b}, characterized by
\begin{equation}
\overline{\Delta \Omega_{\rm d} (t)} = 0, \hspace{3ex}
\text{Var}\{\Delta\Omega_{\rm d} (t)\} = \frac{c \tau}{2} \left( 1- \ee^{-2t/\tau} \right),
\label{eq:dpi1}
\end{equation}
(recall that we set $\Delta \Omega_{\rm d} (0)=0$) and the following  update formula
\begin{equation}
\Delta \Omega_{\rm d} (t + \Delta t) = \Delta \Omega_{\rm d} (t) \ee^{-\Delta t/\tau} + \left[ \frac{c \tau}{2} (1-\ee^{- 2 \Delta t/\tau}) \right]^{1/2} n.
\label{eq:dpi2}
\end{equation}
We expect that typical correlation times for such microwave intensity fluctuations will be much larger than the expected gate times. Accordingly, $\tau=1\,$ms is set to be much larger than the gate time $t_{\rm g}=63\,\mu$s. Moreover, we have assumed relative intensity fluctuations on the
order of $\zeta_{\rm I} = \Delta \Omega_{\rm d}/\Omega_{\rm d}\approx 10^{-4}$, which requires a very accurate stabilization of the microwave sources, but is in principle possible (cite). Identifying the relative fluctuations of the microwave with the standard deviation of the O-U process, the diffusion constant $c$ of 
the O-U process is given by
\begin{equation}
\label{intensity_c}
c=\frac{2  (\zeta_{\rm I}\Omega_{\rm d})^2}{\tau},\hspace{2ex}\zeta_{\rm I}\approx10^{-4}.
\end{equation}
In Fig.~\ref{intensity_noise_process}, we represent the fluctuations of the microwave Rabi frequency given by an O-U process fulfilling the above requirements. The time-scale  in this figure corresponds to a single realization of the driven geometric phase gate, and we typically average over $N_{\rm n}=10^3$ gate realizations.

%%%%%%%%%%%%%%%%%%%%%%%%%%%%%%%%%%%%%%%%%%%%%%%%%%%%%%%%%%%%%%%%%%%%%%%%%%%%%


\begin{references}
\bibitem{note}
Note that some additional differential ac-Stark shifts also arise from the adiabatic elimination in Eq.~\eqref{raman}, but these can be carefully compensated in experiments.
\end{references}
\end{document}